\documentclass[twocolumn]{aastex6} 
\usepackage{morefloats}
\usepackage{natbib}
\usepackage{epstopdf}
\usepackage{graphicx}
\usepackage{hyperref}
\usepackage{float}
\usepackage{color}  
\usepackage[normalem]{ulem}  
\usepackage{bm}
\slugcomment{Accepted by The Astrophysical Journal on 22-11-2016}

\begin{document}

\title{Geometric Corroboration of the Earliest Lensed Galaxy at $\lowercase{z}\simeq 10.8$ from Robust Free-Form Modelling.}
\author{Brian M. Y. Chan\altaffilmark{1,2},
Tom Broadhurst\altaffilmark{3,4},
Jeremy Lim\altaffilmark{1,5},
Jose M. Diego\altaffilmark{6},
Adi Zitrin\altaffilmark{7,8}
Dan Coe\altaffilmark{9},
Holland C. Ford\altaffilmark{10}
}

\altaffiltext{1}{Department of Physics, The University of Hong Kong, Pokfulam Road, Hong Kong}
\altaffiltext{2}{brianchanmy@gmail.com}
\altaffiltext{3}{Department of Theoretical Physics, University of Basque Country UPV/EHU, Bilbao, Spain}
\altaffiltext{4}{IKERBASQUE, Basque Foundation for Science, Bilbao, Spain}
\altaffiltext{5}{Laboratory for Space Research, Faculty of Science, The University of Hong Kong, Pokfulam Road, Hong Kong}
\altaffiltext{6}{IFCA, Instituto de F\'isica de Cantabria (UC-CSIC), Av. de Los Castros s/n, 39005 Santander, Spain}
\altaffiltext{7}{Cahill Center for Astronomy and Astrophysics, California Institute of Technology, MC 249-17, Pasadena, CA 91125, USA; adizitrin@gmail.com}
\altaffiltext{8}{Hubble Fellow}
\altaffiltext{9}{Space Telescope Science Institute, Baltimore, MD, USA}
\altaffiltext{10}{Dept. of Physics and Astronomy, Johns Hokpins university, Baltimore, Maryland, USA}

\setcounter{footnote}{11}

\begin{abstract}

A multiply-lensed galaxy, MACS0647-JD, with a probable photometric redshift of $z\simeq 10.7^{+0.6}_{-0.4}$ is claimed to constitute one of the very earliest known galaxies, formed well before reionization was completed. However, spectral evidence that MACS0647-JD lies at high redshift has proven infeasible and so here we seek an independent lensing based ``geometric redshift'' derived from the angles between the three lensed images of MACS0647-JD, using our free-form mass model (WSLAP+) for the lensing cluster MACSJ0647.7+7015 (at $z=0.591$). Our lens model uses the 9 sets of multiple images, including those of MACS0647-JD, identified by the CLASH survey towards this cluster.  We convincingly exclude the low redshift regime of $z<3$, for which convoluted critical curves are generated by our method, as the solution bends to accommodate the wide angles of MACS0647-JD for this low redshift. Instead, a best fit to all sets of lensed galaxy positions and redshifts provides a geometric redshift of $z\simeq 10.8^{+0.3}_{-0.4}$ for MACS0647-JD, strongly supporting the higher photometric redshift solution. Importantly, we find a tight linear relation between the relative brightnesses of all 9 sets of multiply lensed images and their relative magnifications as predicted by our model. This agreement provides a benchmark for the quality of the lens model, and establishes the robustness of our free-form lensing method for measuring model-independent geometric source distances and for deriving objective central cluster mass distributions. After correcting for its magnification the luminosity of MACS0647-JD remains relatively high at $M_{UV}=-19.4$, which is within a factor of a few in flux of some surprisingly luminous $z\simeq 10$--$11$ candidates discovered recently in Hubble blank field surveys.
\keywords{gravitational lensing: strong —-- galaxies: high-redshift —-- (cosmology:) dark matter}


\end{abstract}

\section{Introduction}\label{sec:intro}

Cluster lensing has the advantage over field surveys of a magnification boost and also the ability to measure distances ``geometrically'' from the angular separations within each set of multiply-lensed images \citep{broadhurst05, zitrin14}. This ability provides a welcomed check of photometrically derived redshifts, particularly at high redshift where detections are weakest and restricted to fewer passbands, and for which a dusty-red galaxy solution at much lower redshift is usually feasible. For this purpose an accurate lens model is required, ideally based on many sets of multiply-lensed images spread over the full range of source distances. Since the angles through which light is deflected scale with increasing source distance behind a given lens, we can convert these distances to source redshifts for a given cosmology. This ``geometric'' redshift can then be compared with independently derived photometric redshifts to discriminate between degenerate photometric redshift possibilities. This method has been applied to the large lensing cluster A1689 where geometric distances provided a consistency check of the lens model \citep{broadhurst05, diego14, limousin07}, and for the recently claimed high redshift of a triple-lensed image at a photometric redshift of $z\simeq 9.6$ \citep{zitrin14} behind A2744 in deep HFF data.

Currently, efficient detection of high redshift galaxies is best achieved using the IR channel of the HST Wide-Field Camera-3 (WFC3), supported by the Spitzer Space Telescope's Infrared Array Camera (IRAC) in the mid-IR.  This combination is now generating statistically useful samples of dropout galaxies to $z\simeq 8$ in several independent deep field surveys \citep{oesch10, oesch13, ellis13, finkelstein13, holwerda15, schmidt14}. Examples include high redshift galaxies from the CANDELS \citep{grogin11, koekemoer11} survey where relatively strong [OIII] and $H\beta$ emission creates a mid-IR excess to provide the highest redshift spectroscopically confirmed galaxies \citep{oesch15} reaching $z=8.68$ \citep{zitrin15b}, with confirming strong  Ly-$\alpha$ emission from Keck IR-spectroscopy. Lensing by galaxy clusters has provided a number of even higher redshift candidates, with the highest redshift lensed galaxy being a triply-lensed small round object MACS0647-JD that lies an estimated photometric redshift of $z\simeq10.7^{+0.6}_{-0.4}$ \citep{coe13}, discovered behind the massive lensing cluster MACSJ0647.7+7015\citep{ebeling07,zitrin11},
followed by a similar object at $z=9.6\pm0.2$ \citep{zheng12} behind the CLASH cluster MACS1149+22. Deeper Hubble Frontier Field imaging (HFF), with 70 orbits of optical/IR imaging per cluster, has revealed a $z\sim10.1$ candidate \citep{infante15} in the cluster MACS J0416-2403, likely to be one the faintest galaxies to date. In addition, another triply-imaged galaxy has been uncovered by the HFF at $z\sim 9.8\pm0.3$, magnified by a factor of ~20, by the lensing cluster A2744, that is supported geometrically in \citep{zitrin14} with our free-form model.

Deep Hubble grism observations by \citet{pirzkal15} of MACS0647-JD do not detect emission lines at $z\simeq 2$, adding weight to the higher redshift interpretation. Hubble grism observations have also detected the Lyman break of a relatively bright galaxy GD11.1 at $z\simeq11.09^{+0.08}_{-0.12}$ by \citet{oesch16}. This may strengthen the case for several other relatively bright and high redshift galaxies claimed in the range $z\simeq$ 10--11, in the combined deep field analysis of \citet{holwerda15}, from which GD11.1 (originally called GN-z10-1 in \citet{holwerda15} had an estimated photometric redshift of  $z=10.1$) was identified. Placing this object at $z=11.1$ implies an exceptional brightness of  $M_{UV}= -22.1\pm0.2$ and this object is marginally resolved, so dominant AGN emission can be excluded. The luminosities of all 6 of the \citep{holwerda15} objects are exceptionally high, and it is interesting that there seem to be an absence of lower luminosity galaxies at this redshift that could have been detected \citep{oesch16}. This tendency towards unexpectedly high luminosity may conceivably indicate that $z\sim11$ represents the earliest limiting time that galaxies formed, with nothing detected beyond. This conclusion is provisional at the moment given HST's limit is effectively not far above $z \simeq 12$ for detecting Lyman Break Galaxies (LBGs) in the longest wavelength 1.6$\mu$m passband. 

The improving quality of lensing data for clusters imaged with Hubble and the future with JWST and EUCLID has encouraged us to focus on  understanding and improving our free-form lensing method \citep{diego05b, sendra14}. We have applied this method to all of the HFF fields and some CLASH clusters, taking advantage of the increasing numbers of multiply lensed images and their photometric and spectroscopic redshifts as the input data \citep{lam14,diego15a,diego15b,diego16a,diego16b}. We have made these models available to the community so that they may be compared to those derived by other teams.  Parametric models may be regarded as best suited to virialized clusters \citep{halkola06, limousin07} for which substructure is minimal but may be extended to accommodate obvious bimodal substructure, and confirmed in blind tests on an idealised bimodal model cluster \citep{meneghetti16}. In general, however, the complexities of massive merging clusters such as those chosen for the HFF program require the definition of several new parameters for each additional model halo, where the definition of substructure and its location is less than objective leading to ambiguity. It is important to have free-form capability even when modelling clusters with relaxed appearance as these can contain levels of structure in detail  locally affecting the position and brightness of individual lensed images, as we show here in this work and also in blind lensing comparison tests made by \citep{meneghetti16} where a relaxed, simulated cluster with s typical level of substructure was found to be very well reproduced by the free form methods, including WSLAP+ that we use here. Free form modelling may be expected to excel relative to parametric models for the largest lenses, which are typically clusters in a state of collision with convoluted critical curves, including most of the HFF clusters for which WSLAP+ has proven capable of locating new sets of multiple images by virtue of its flexibility \citep{diego14,diego16b,lam14}, and also in making successful predictions for the return of the ``Refsdal'' multiply-lensed SN in HFF data \citep{diego16a,treu16} and for the intrinsic luminosity of a magnified SN1a in the HFF cluster A2744 \citep{rodney15}.

The extent to which parametric lens models capture substructure and tidal distortions of the dark matter distribution, including relatively relaxed and merging clusters, has prompted us to look harder at the possibility of grid solutions, where the lens plane is represented by a uniform or an adaptive grid of Gaussian pixels. Early non-parametric studies with a simple uniform gridding of the mass distribution were not accurate enough for identifying new sets of multiple images because they did not have high enough resolution to capture the local perturbing effects of cluster member galaxies. Typically at least one member of any set of multiply lensed images will fall close to a member galaxy, with additional images often created in this way, for which the limited spatial resolution of a smooth grid does not capture. We have achieved a huge improvement recently in this approach by incorporating a simple model halo for each of the observed member galaxies, together with the smooth grid to model more distributed cluster mass in a uniform grid \citep{sendra14}. Our simulations and successful applications of this improved free-form model have demonstrated that meaningful solutions to be found as the small scale deflections and additional multiple images locally generated by the member galaxies can be accounted for and new sets of multiple images can be identified.

We have demonstrated now with our HFF work that this free-form approach generates lens models that are sufficiently accurate to predict the locations of counter-images \citep{lam14} and can correct and complete ambiguous identifications reported in competing work, so that physically plausible mass distributions can be derived that are relatively free of model assumptions.  The reliability of our method has been demonstrated with both simulated data \citep{sendra14} and observations of the lensing standard Abell 1689, \citep{diego14} and our subsequent HFF work \citep{lam14, diego15a, diego15b, diego16b, diego16a}. In particular we have stressed the ability of our models to account for the independent information contained in the relative fluxes of multiply lensed sources. We have shown that our models find a  linear relationship between the predicted model magnifications and the observed fluxes for the multiply-lensed sources 
\citep{lam14} lying symmetrically about the one of equality, with a dispersion consistent with the observed errors. Here we show that this constancy also holds here for our new free-form model of MACS0647+7015, adding great confidence in the objectivity of our free-form modelling of this cluster.

Previous lens models and input data for our lens model of MACS0647 are summarised in section \ref{sec:previous}. The image and photometric redshift source is presented in section \ref{sec:color}, as well as detailed description of each of the 9 lensed image systems in turn. In section \ref{sec:model} we describe our methodology of constructing a free-form lens model. We define ``geometric redshifts" in section \ref{sec:geoz}, based on the distance-redshift scaling for multiply-lensed galaxies. Our main results are described in section \ref{sec:results}, with a demonstration on the predictive power on magnifications in section \ref{sec:mag_prediction}. The consistency check on geometric and photometric redshift comparisons are presented in section \ref{sec:geoz_consistency}. We include the model prediction on extra lensed images in section \ref{sec:other_images}, with discussions and conclusions in section \ref{sec:discussion} and \ref{sec:conclusions} respectively. The mass model uncertainties are presented in the Appendix. Standard cosmological parameters are adopted: $\Omega_M=0.3, \Omega_\Lambda=0.7$ and $h=H_0/100\,{\rm km\,s^{-1}\,Mpc^{-1}}=0.7$.

\section{Available lensing data and previous lens models}\label{sec:previous}
  
Here we rely on the CLASH survey for our input data when modelling MACSJ0647.7+7017. These data include the reduced images, photometry, BPZ based photometric redshifts derived by the method of \citet{coe13}, and positions of lensed images in \citet{coe13} and \citet{zitrin11,zitrin15a}. There are a total of 9 multiply lensed systems forming 24 identified images spread uniformly over the critical area of the cluster. The faintest image detected is around an AB magnitude of 28.5. Details of all images are listed in Table \ref{t:image_data}.  There are three previous lens models applied in \citet{coe13} for modelling this cluster, constituting both parametric and free-form methods. The ``light approximately traces mass" method of \citep{zitrin09a,zitrin09b,zitrin11} has been proven successful in identifying over one thousand multiply-lensed images in the CLASH survey and beyond. This method may be described as ``semi-parametric" in that it is based initially on a superposition of dark matter for the visible cluster member galaxies and a smooth component is fitted to this with a general low order polynomial fit to allow flexibility in the unknown dominant smooth cluster component. The member galaxies are then modelled by \citet{zitrin11} for MACS0647 simply as the residual of this cluster component after subtracting from the initial member galaxy composite, providing a minimalistic method that includes member galaxies and allows for some flexibility in the definition of a smooth cluster wide mass distribution following \citet{broadhurst05}. This method was developed to take advantage of the approximate relation expected between dark matter and the distribution of member galaxies that is generally close but not expected to be exact, due to tidal forces shaping the DM and the finite numbers of tracer galaxies that can be utilised.
 
\citet{coe13} also make use of Lenstool \citep{kneib11} - the most successful fully parametric method which is best suited to symmetric lenses where the definition of the cluster mass distribution is not very ambiguous. In the case of merging asymmetric clusters there is the inherent problem with this tool of deciding how to deal with mass substructures
within the cluster dark matter -  for each substructure that is introduced, at least 6 parameters must be added to 
the model including centroid, ellipticity, position angle, profile gradient and amplitude, with an inherent restriction 
to a given class of such profiles.  This model has been 
relied on for estimating the inherent magnification of the lensed sources, including the three counter-images of the object of interest here MACSJ0647-JD.
   
Below we describe in greater detail the very general Free-Form lensing method that we have increasingly employed in our work on the HFF. This method takes advantage of the high density of multiply-lensed images to provide sufficiently constraints on the smooth cluster mass distribution. The robustness and predictive power of the free-form method have been demonstrated by the capability we have to predict counter-images in the HFF data, and in some cases correcting the parametric model results of Lenstool in the case of the complex cluster A2744 \citep{lam14}. Furthermore, because this method is largely model free, the geometric distances we derive are objective and not tied to any assumed  cluster mass profile.

\section{Colour Images and Photometric Redshifts}\label{sec:color}


The retrieved drizzle science images were obtained from the online CLASH archive.  
The individually reduced images are produced by \textit{Mosaicdrizzle} \citep{koekemoer13} and publicly available\footnote{https://archive.stsci.edu/missions/hlsp/clash/macs0647/\\data/hst/scale\_65mas/}. Each cluster was observed with a a wide filter coverage from 225nm to 1600nm. The images can probe faint galaxies to 
exceptional depth, and with their filter coverage photometrically identify galaxy candidates when the Universe was younger than 800 million years old - or less than 6\% of its 
current age.

\begin{figure}
\includegraphics[width=85mm]{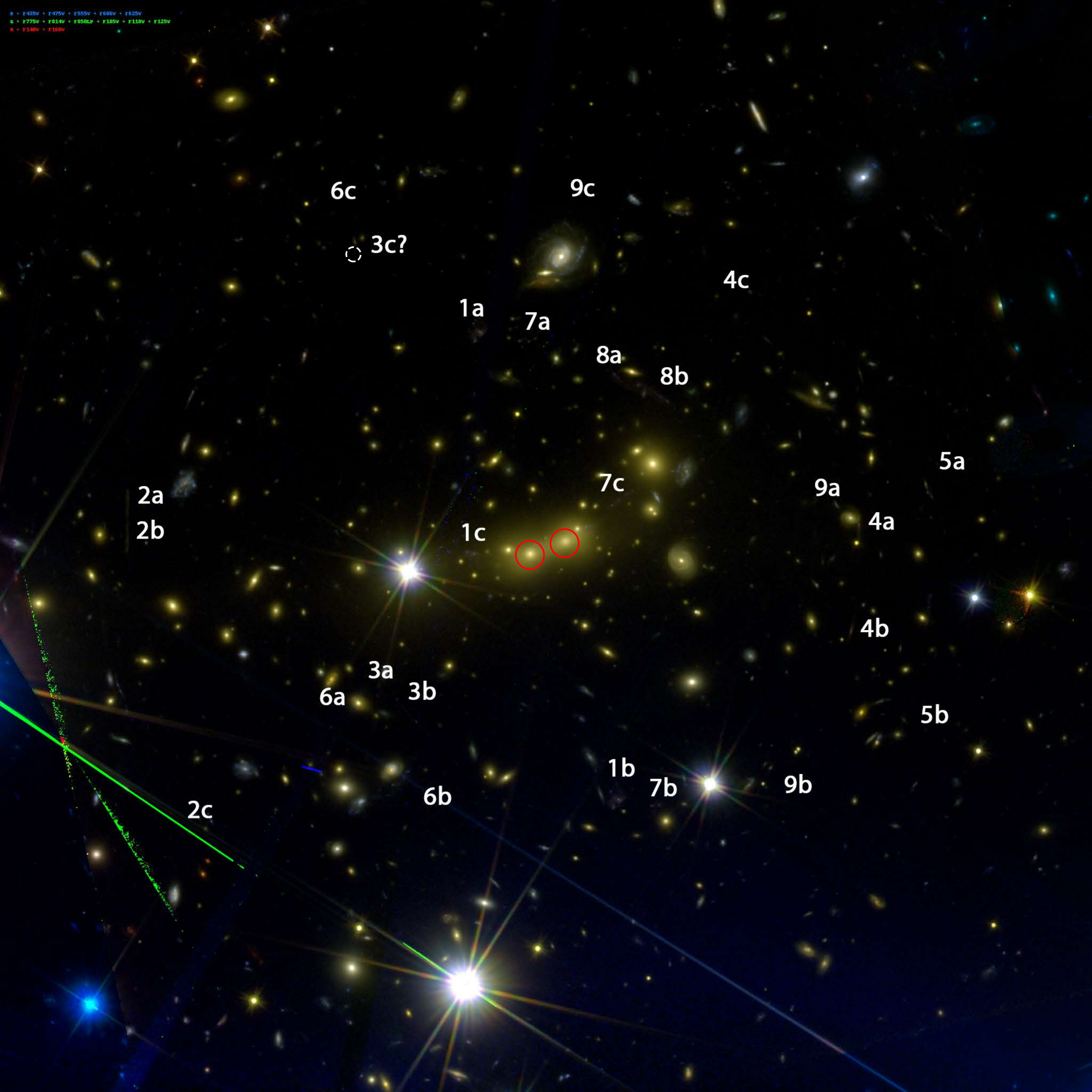}
\caption{
ID's of galaxies lensed to form multiple images by MACS0647+7015. In the process of generating this image, F160W and F140W construct the red colour, F125W, F110W F105W, F850lp, F814W, and F775W contribute to green colour, F625W, F606W, F555W, F475W, and F435W are used in blue colour.
The field of view is 133.12"$\times$133.12". 
The separately modelled BCGs are circled in red.
}
\label{f:combo_map}
\end{figure}
We generated a simple RGB image (Figure \ref{f:combo_map}) by adding all filters that have been assigned to R,G and B colours separately. The added images are combined in the publicly available software \textit{Trilogy}\footnote{http://www.stsci.edu/$\sim$dcoe/trilogy/Intro.html}.

The lensed image positions and redshifts were previously presented by \citet{coe12}. In \citet{zitrin15a}, three more candidate systems were identified, referred to henceforth as systems 10 to 12. These three sources were not included in our mass model as we found the self-consistency of our mass model to be generally reduced after the addition. This reduction of self-consistency can be explained by either false identification of lensed images or insecure photometric redshifts. When we analyzed the self-consistency of the mass models, we found 9 systems to be already sufficient in constraining the mass model to a level of precision that achieves excellent self-consistency. 

We conducted our own photometry for each individual candidate lensed galaxy. Tailor-made apertures and sky annulus were applied to each multiply lensed images in a combined image of all WFC3 infra-red bands (consisting of F105W, F110W, F125W, F140W and F160W). We performed the photometry on a combined image of multiple bands as most lensed images are very faint. When the photometry is conducted on faint images, we used apertures approximately equal to Rayleigh diffraction limit of the Hubble Space Telescope that give us the highest S/N ratio.

\section{Free-Form Lensing Model}\label{sec:model}

The free form lensing method developed by \citet{diego05a} is a grid-based iterative method that can be constrained by both strong and weak lensing information, including sets of individual pixels subtended by resolved arcs in the case of strong lensing. 
We have recently demonstrated that this method can be significantly improved by the addition of observed member galaxy deflections (Weak and Strong Lensing Analysis Package plus member galaxies: WSLAP+, \citet{sendra14}) because typically one or more counter-images of each multiply-lensed system is either generated or significantly deflected by a local member galaxy.  We have applied this method recently to the relaxed cluster A1689 \citep{diego14}, the HFF clusters A2744 \citep{lam14}, and MACS0416 \citep{diego15a}. We have demonstrated that this combination of high and low frequency components can converge to meaningful solutions with sufficient accuracy to allow the detection of new counter-images for further constraining the lensing solution of A1689. This free-form method is especially useful for modeling the complex mass distributions of the merging clusters chosen for the Hubble Frontier Fields program. Parameterised models are inherently less useful in this context as the number of constraints is generally insufficint to model the additional parameters. We also apply this free-form method here as the structure of MACS0647+7015 is very elongated with some level of substructure anticipated and also because we wish to obtain relatively model-independent geometric redshift estimates for
the lensed galaxies, particularly for MACS-JD, free from any restrictions inherent to idealised parametric profiles.

Here we outline briefly WSLAP+ for the mass reconstruction and refer the reader for details of its implementation in our previous papers\citep{diego05a, diego05b, diego07, ponente11, sendra14, diego14}. \\

\noindent Given the standard lens equation, 
\begin{equation}\label{eq_lens}
\bm{\beta} = \bm{\theta} - \bm{\alpha}(\bm{\theta},\Sigma(\bm{\theta})),  
\end{equation} 
where $\bm{\theta}$ is the observed angular position of the source, $\bm{\alpha}$ is the deflection angle, $\Sigma(\bm{\theta})$ is the surface mass density of 
the cluster at the position $\bm{\theta}$, and $\bm{\beta}$ is the position of the background source, both the strong lensing and weak lensing observables can be expressed in terms of derivatives of the lensing potential
\begin{equation}\label{eq_psi} 
\psi(\theta) = \frac{4 G D_{l}D_{ls}}{c^2 D_{s}} \int d^2\theta'
\Sigma(\theta')ln(|\theta - \theta'|), 
\end{equation}
where $D_l$, $D_{ls}$ and $D_s$ are, respectively, the angular diameter distances to the lens, from the lens to the source, and from the observer to the source. 
The unknowns of the lensing problem are in general the surface mass density and the positions of the background sources. 
As shown in \citet{diego05a}, the lensing problem can be expressed as a system of linear equations that can be represented in a compact form,
\begin{equation}\label{eq_lens_system} 
\Theta = \Gamma X,
\end{equation} 
where the measured lensing observables are contained in the array $\Theta$ of dimension $N_{\Theta }=2N_{SL}$, the unknown surface mass density and source positions are in the array $X$ of dimension $N_X=N_c + N_g + 2N_s$, and the matrix $\Gamma$ is known (for a given grid configuration and initial galaxy deflection field, see below) and has dimension $N_{\Theta }\times N_X$.  
$N_{SL}$ is the number of strong lensing observables (each one contributing with two constraints, $x$, and $y$), and $N_c$ is the number of grid points (or cells) that we divide the field of view (133.12"$\times$133.12") into, which equals to $32^{2}=1024$ in this case.  
$N_g$ is the number of deflection fields (from cluster members) that we consider.  
$N_s$ is the number of background sources (each contributes with two unknowns, $\beta_x$, and $\beta_y$, see \cite{sendra14} for details). 
The unknowns are found after minimizing a quadratic function that estimates the solution of the system of equations \ref{eq_lens_system}, with the constraint that the solution, $X$, must be positive, and is constrained not to generate sources that are unreasonably small given the angular sizes of faint galaxies 
resolved by Hubble.  These constraints are particularly important to avoid the unphysical situation where the masses associated to the galaxies are negative (which could otherwise provide a reasonable solution, from the formal mathematical point of view, to the system of linear equations \ref{eq_lens_system}). 
Imposing the constraint $X>0$ also helps in regularizing the solution as it avoids large negative and positive contiguous fluctuations.

For the model galaxy component we use the set of cluster member galaxies as used in \citet{zitrin15a}. These galaxies are selected first from the red sequence, and then vetted by eye with a few obvious non-members excluded. From the H band (F160W) magnitudes, a mass-to-light ratio of 20 M$_{\odot}$/L$_{\odot}$ is initially assumed to construct the fiducial deflection field summed over the member galaxies, each having a truncated NFW profile (truncation radius equals scale radius times concentration parameter) with a scale radius linearly related to its FWHM in the NIR image. 
For our purpose, the exact choice of profile for member galaxies is not particularly important; what matters more is the normalisation. 
This normalization is the only free parameter of the fiducial deflection field, and is determined by our optimization procedure. 
In \cite{sendra14} we tested this addition to the method with simulated lensed images, but with the real galaxy members from A1689 to be as realistic as possible. 
We found that a separate treatment of the BCG lensing amplitude was warranted, adding a second deflection field i.e $N_g=2$ (see definition of $N_g$ above) to be solved for. 
Here we follow the same procedure as A2744 incorporating member galaxies and the BCGs separately. 
We also find significant improvements in the residual by leaving free the amplitude of bright galaxies that significantly perturb nearby lensed images.  In total, we decomposed the fiducial deflection field into 7 components, comprising the two central galaxies, 5 components of member galaxies near to different lensed images, and one component for all other member galaxies.

\section{Geometric Redshifts}\label{sec:geoz}

The derived distances provide a very welcome check on redshifts derived photometrically, particularly at high redshifts where images are generally noisy and may be detected in only the longest-wavelength passband.  For this purpose an accurate lens model is required, based on many sets of multiply-lensed images and ideally sampling a wide range of source distances so that the gradient of the mass profile can be constrained. The reduced deflection field scales with increasing source distance behind a given lens so that the separations in angle between images of the same system are larger for higher redshift sources. 
Therefore, we have developed a code that can ``delens" any particular image to the source plane, and ``relens" it back to the image plane. We repeat the procedures for a redshift range around the claimed redshift of the source. This procedure defines ``loci" along which we search for counter images. By comparing the loci with the observed positions of counter images, we obtain geometric distance estimates for each set of multiple images. Distances derived this way can then be converted via cosmological parameters to source redshifts and compared with independently derived photometric redshifts.  This method has been established using the large lensing cluster A1689, where geometric distances provided a consistency check of the lens model \citep{broadhurst05, limousin07, diego14, lam14}.

A lens model is a deflection field, $\bm{\alpha}_L(\bm{\theta})$, that expresses the angle through which light is bent at the lens plane. 
An observer sees a reduced angle scaled by a ratio involving lens and source distances:

\begin{equation}\label{eq_reduced_angle}
\bm{\alpha}(\bm{\theta})=\frac{d_{ls}(z)}{d_{s}(z)} \bm{\alpha}_{L}(\bm{\theta})
\end{equation}

\noindent As it can be observed from equation \ref{eq_reduced_angle}, the angles between the unlensed source and the lensed images slowly increase with source distance behind  a given lens. This dependence means that the locations of a given set of multiple images will meet most closely in the source plane at a preferred source distance. In principle, we can only determine relative distances this way because the absolute value of $\bm{\alpha}_L(\bm{\theta})$ cannot be determined independently of lensing.  By normalizing the model deflection field using a spectroscopic redshift measured for any one of the multiply-lensed systems, the relative distances can be converted to absolute distances for a given cosmological model. In other words, what we actually determine, for the $k^{th}$ set of multiple images for a given lens, is the ratio of lensing distances:

\begin{equation}
f_k(z)={d_{ls_k}(z) \over d_{s_k}(z)} / {d_{l{s_o}}(z_o) \over d_{s_o}(z_o)}
\end{equation}
where the $o$ index refers to the multiply-lensed system with spectroscopic redshift.

Consider an image $i$ from source $k$, we can write the lens equation as:

\begin{equation}
	\bm{\beta}_k = \bm{\theta}_{k,i} - \frac{d_{ls_k}(z)}{d_{s_k}(z)} \bm{\alpha}_{L}(\bm{\theta}_{k,i}), 
\end{equation}

Taking the difference of equation 6 from image $i$ and $j$, we could eventually get this expression:

\begin{equation}
	\frac{d_{ls_k}}{d_{s_k}} = \frac{(\bm{\theta}_{k,i} - \bm{\theta}_{k,j}) \cdot (\bm{\alpha}_L(\bm{\theta}_{k,i}) - \bm{\alpha}_L(\bm{\theta}_{k,j}))}{[\bm{\alpha}_L(\bm{\theta}_{k,i}) - \bm{\alpha}_L(\bm{\theta}_{k,j})]^2}
\end{equation}
 
We are free to normalise the $f_k$ ratios as convenient, and this is sensibly done relative to a multiply lensed source with a reliable spectroscopic redshift, which is not available for this cluster, so we  adopt the maximum value of $d_{ls}/d_s$ for a source at infinity, and with the lens redshift equal to that of the cluster $z=0.591$. For a system involving multiple images, we can calculate the statistical mean of all the combinations of image $i$,$j$:

\begin{equation}
	f_k = \frac{2}{n_k(n_k-1)}\sum\limits_{i,j,i>j}\frac{(\bm{\theta}_{k,i} - \bm{\theta}_{k,j}) \cdot (\bm{\alpha}_L(\bm{\theta}_{k,i}) - \bm{\alpha}_L(\bm{\theta}_{k,j}))}{[\bm{\alpha}_L(\bm{\theta}_{k,i}) - \bm{\alpha}_L(\bm{\theta}_{k,j})]^2}
\label{eq_f_k_equation}
\end{equation}
In this expression, we compare the right hand side expression, which depends on the ``goodness" of our model, with the theoretical $d_{ls}(z)/d_{s}(z)$ curve to evaluate the reliability of our lens model. Below we present a comparison of $f_k$ values in mass models generated by assuming different redshifts of MACS0647-JD.

\section{Results}\label{sec:results}

\subsection{Low Redshift Solution}
\begin{figure}
\includegraphics[width=85mm]{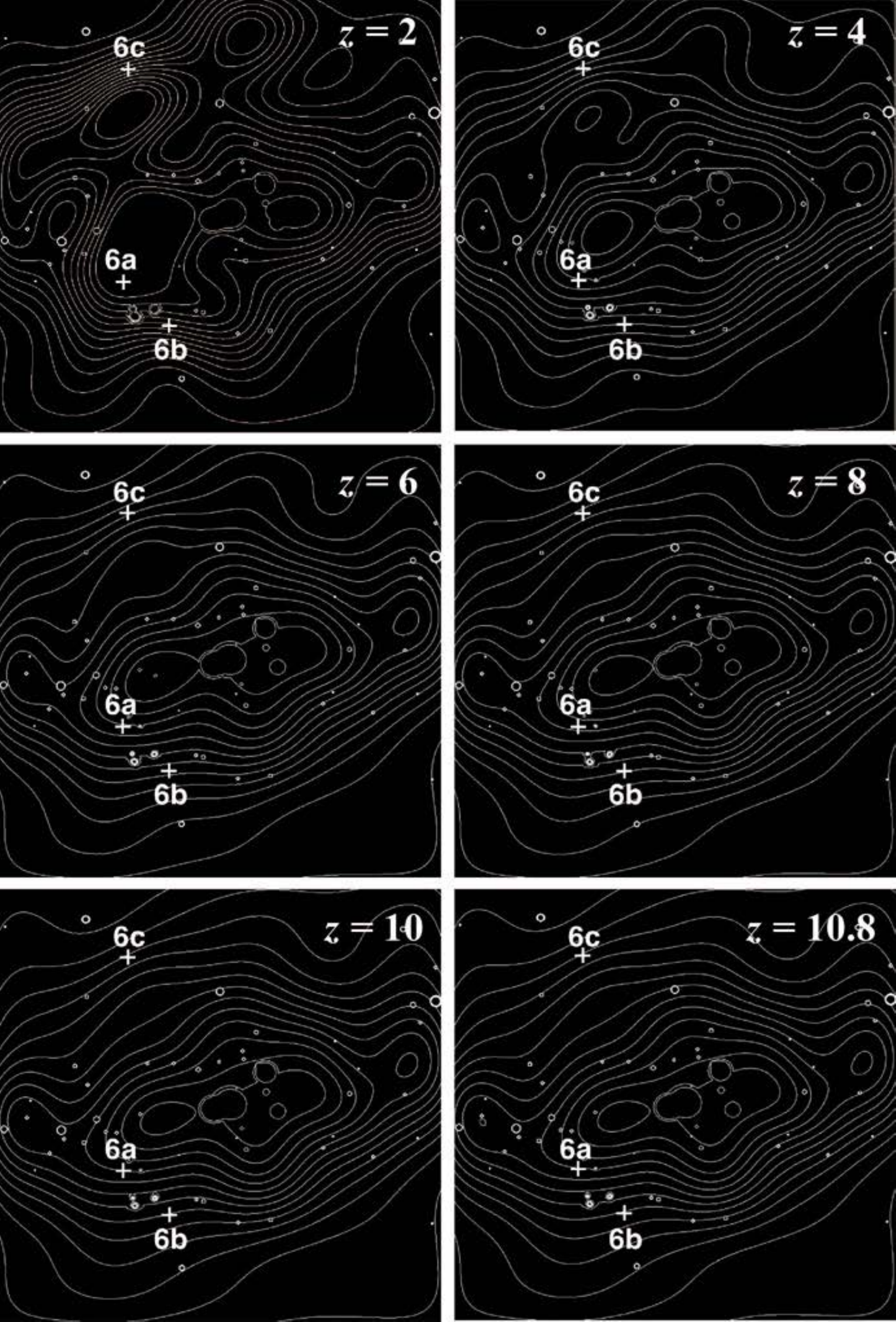}
\caption{
Projected mass contours of our lens model generated for a range of different redshifts of system 6, from $z=2$ to $z=10.8$. The contours are spaced linearly in 14 steps. This shows how the model requires a significantly more structured mass distribution for a lower choice of redshift for system 6, particularly near the outlying image 6c, becoming generally much smoother for a choice of  $z=11$ for system 6. Note that in making these models all other 8 multiply lensed systems at lower redshift are placed at their best photometric redshifts - ranging from $1.9<z<6.5$, with only system 6 varied in redshift between these plots.
}
\label{f:mass_contours}
\end{figure}

\begin{figure}
\includegraphics[width=85mm]{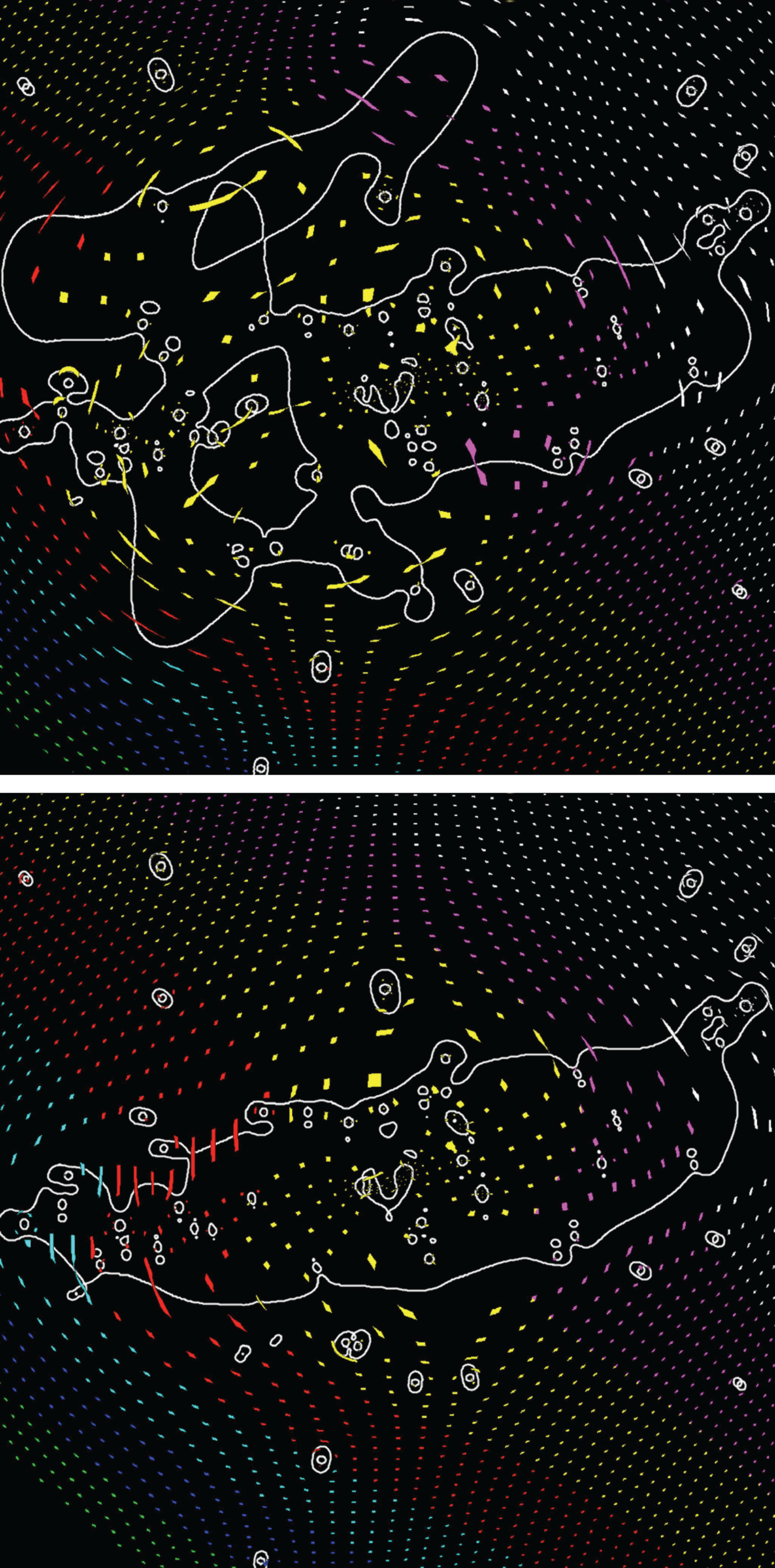}
\caption{
The critical curves generated when assuming a low redshift, $z=2$, for system 6 (MACS0647-JD) are shown in the upper panel, and the same plot assuming $z=11$ in lower panel.  These critical curves cross each other in multiple places unlike the observed critical curves that follow a much simpler elliptical pattern like the distribution of the member galaxies and are delineated clearly by the observed close pairs of multiple images and large arcs (see Figure \ref{f:combo_map}). 
these critical curves are overlaid on a plane of small evenly spaced sources to illustrate further the large differences between the low redshift $z=2$ solution (top) and the best fit model with system 6 at $z=11$ with colour coded diagonal stripes linking related multiple images. It is impressive that the free form approach is capable of testing this low redshift, providing a clear negative answer.
}
\label{f:weird_curve}
\end{figure}

To examine the geometric constraints on the redshift of JD, we reconstructed the mass model of the cluster using all the input data in Table \ref{t:image_data} but allowing for a wide range of input redshift for 
system 6. The results are shown in Figure \ref{f:mass_contours} with assumed redshifts for JD of $z=$ 2, 4, 6, 8, 10, and 10.8.  It is clear from this that the free form model is capable of great versatility in fitting all the multiple image data, as can be seen in Figure \ref{f:weird_curve} where we plot the critical curve for a lensed system at $z=3$ by setting $z=2$ for system 6. With this low redshift, the lensing mass contours of the model greatly deviate from the observed distribution of central cluster members, with correspondingly very convoluted critical curves. The critical curves shown in Figure \ref{f:weird_curve} can be completely excluded as they cross each other in multiple places. It is very clear where the critical curves lie from the pattern of observed close pairs of multiple images and large arcs, matching closely the elongated distribution of member galaxies (Figure \ref{f:combo_map}), apparent in the higher redshift solutions in Figure \ref{f:mass_contours}.

\begin{figure}
\includegraphics[width=85mm]{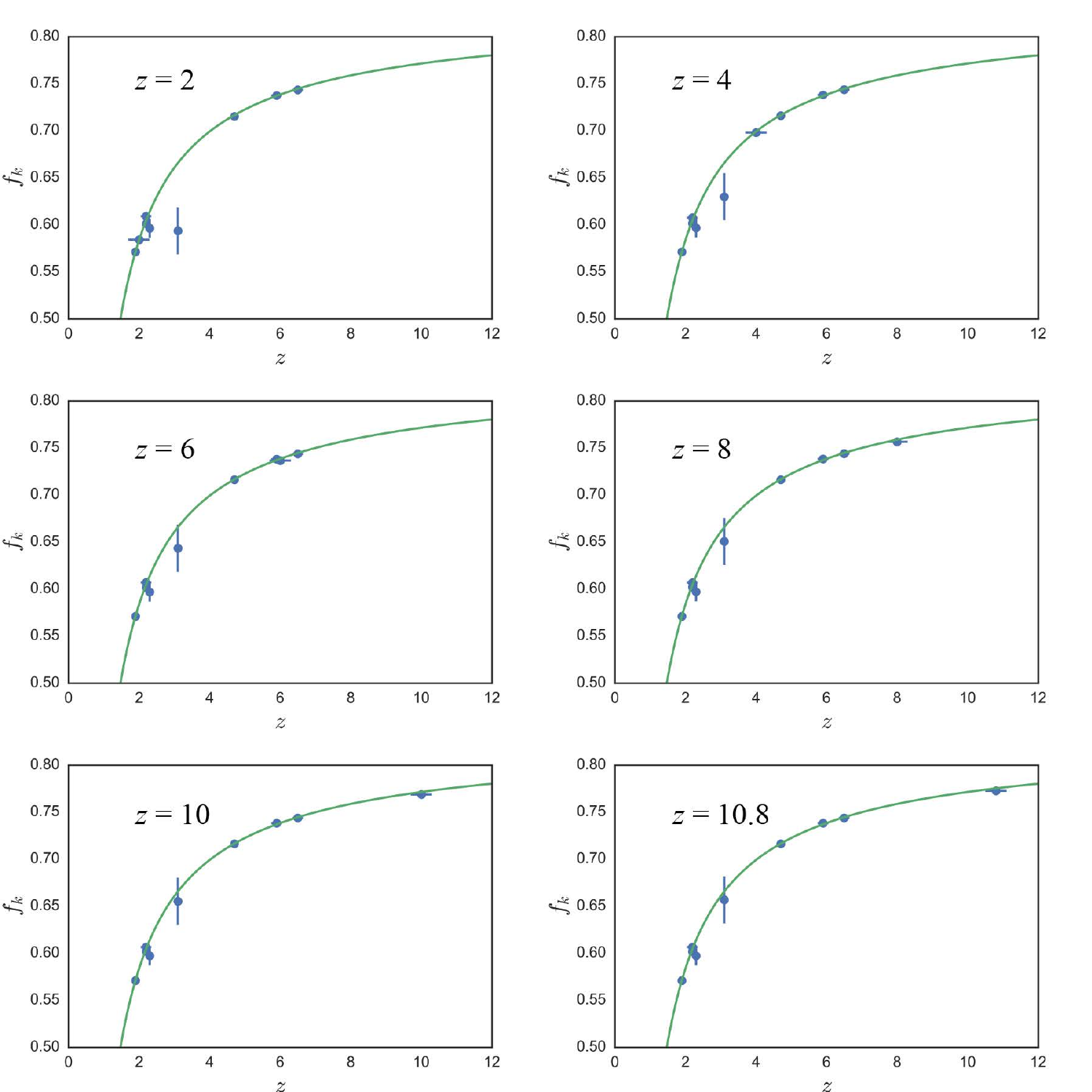}
\caption{
	The lensing distant ratio, $f_k$ (green dots) derived by the model for each multiply lensed system plotted against their photometric redshifts, where we explore a range of redshift for system 6, as indicated on each panel. The blue line shows the theoretical $d_{ls}(z)/d_{s}(z)$ curve, for the standard cosmology which is very well fitted by the data in general, especially for a high redshift of system 6. The sensitivity of this fit is strongest for system 3 which lies relatively close to MACS0647-JD and improves as the redshift of MACS0647-JD is increased. Error bars of the y-axis represents the standard deviation of 46 model reconstructions each with different grid resolutions, initial conditions and photometric redshifts assumptions (see Appendix). Error bars of the x-axis represents 2$\sigma$ dispersion of photometric redshifts as presented in \citet{coe13}.
}
\label{f:f_k}
\end{figure}

\begin{figure}
\centering
\includegraphics[width=75mm]{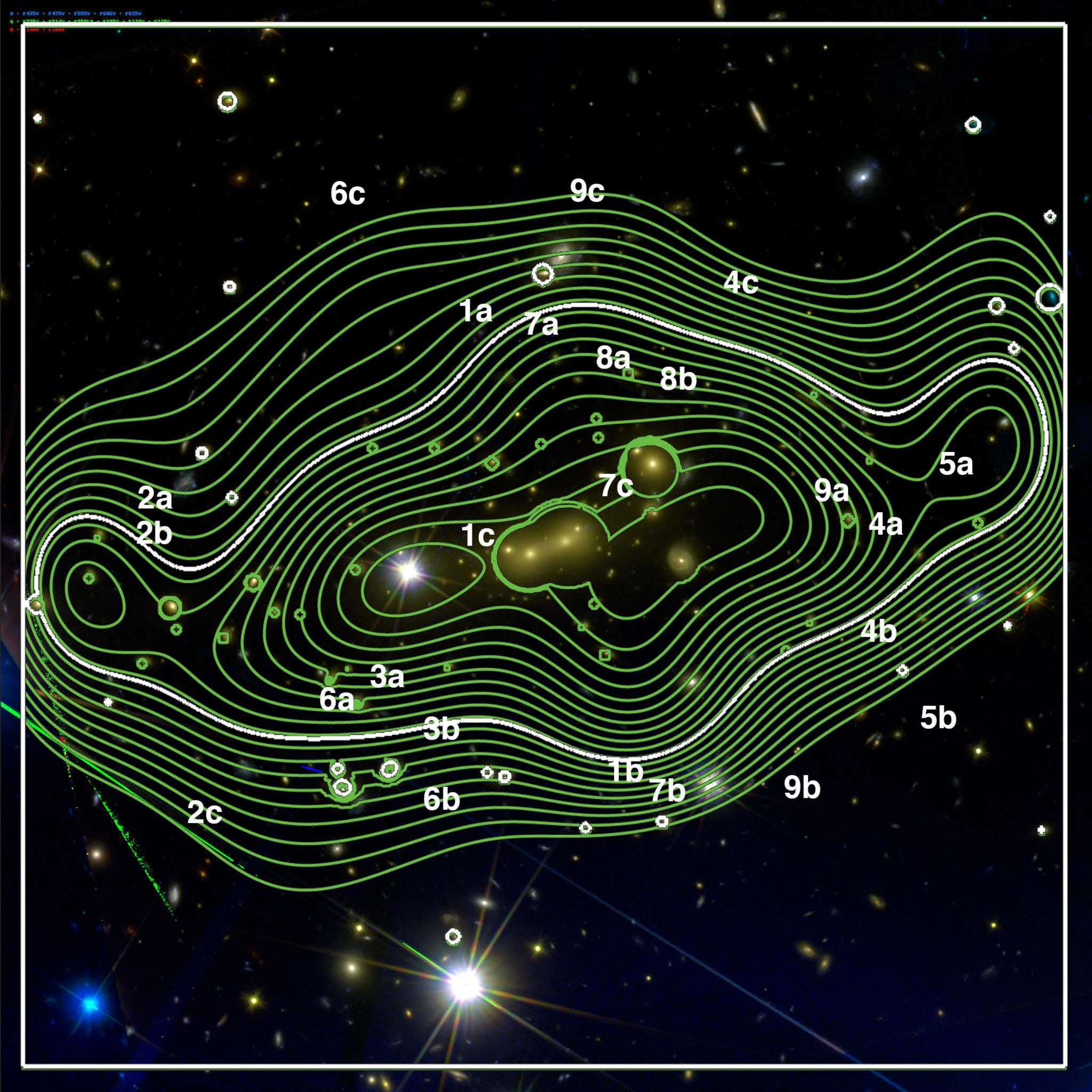}
\caption{
Linearly spaced contours of our lens model overlaid on our colour image of MACS0647. 
The green contours spread from $5.04 \times 10^{14} M_{\odot}Mpc^{-2}$ to $2.75 \times 10^{15} M_{\odot}Mpc^{-2}$ with 30 contour lines. The half critical density contour at $z=3$ is coloured in white. 
The field of view is 133.12"$\times$133.12". 
}
\label{f:mass}
\end{figure}

\begin{figure}
\centering
\includegraphics[width=75mm]{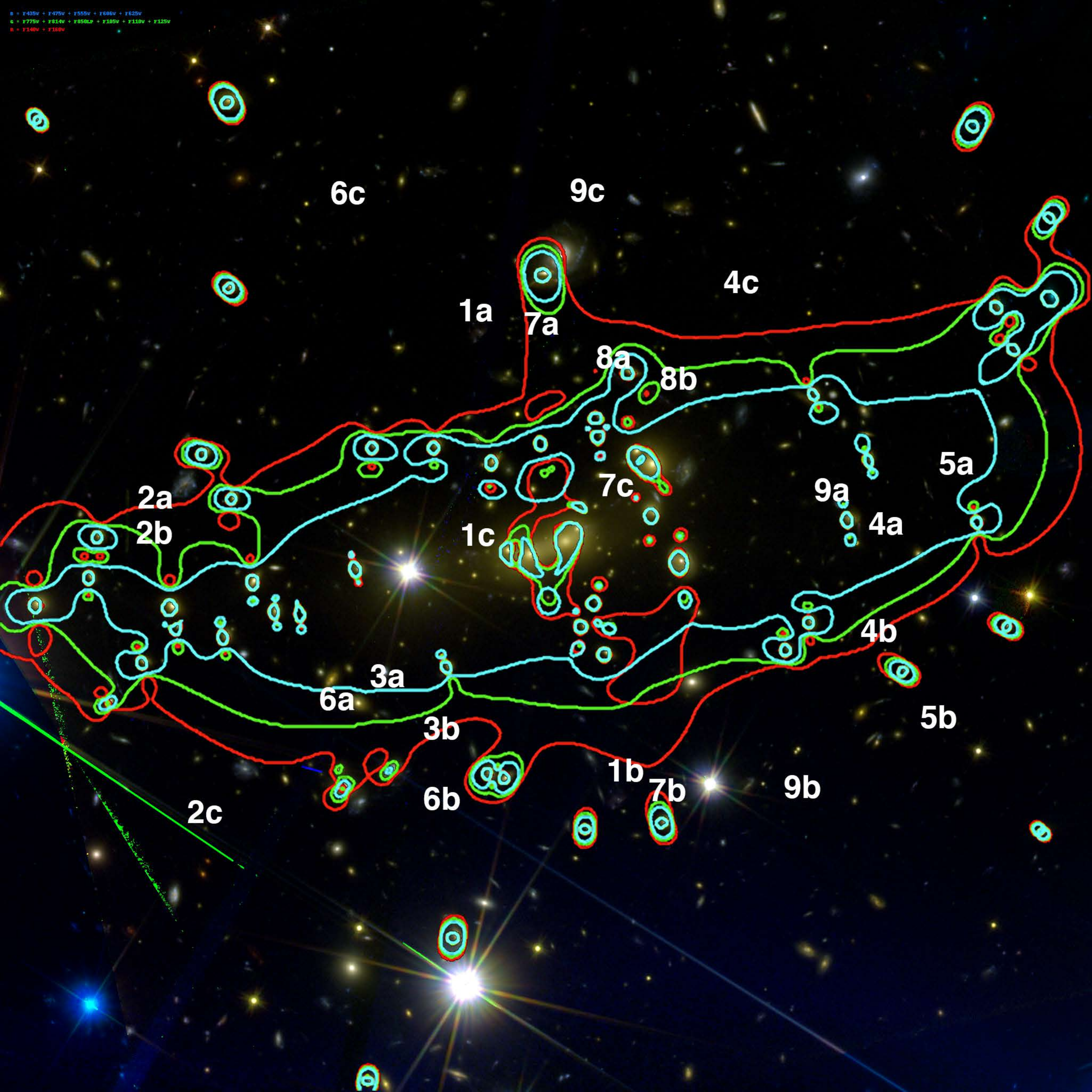}
\caption{
The critical curves at $z=2$ (cyan), $z=3.5$ (green), $z=11$ (red).
The field of view is 133.12"$\times$133.12". 
}
\label{f:critcurve1}
\end{figure}

\begin{figure}
\centering
\includegraphics[width=75mm]{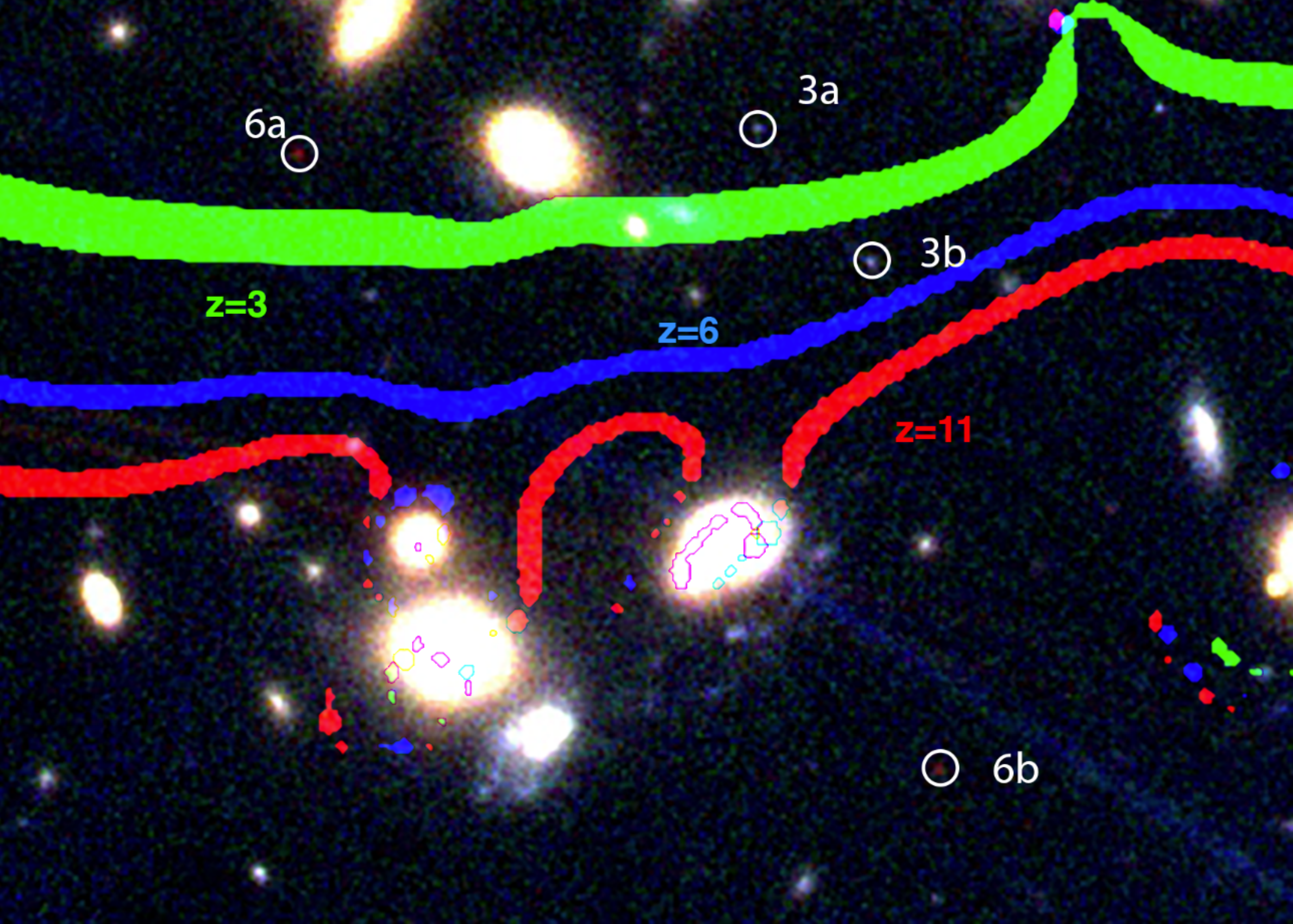}
\caption{
A close-up on system 3 and 6 with critical curves at $z=3,6,11$. The contrast has been adjusted to better distinguish the lensed images.
}
\label{f:critcurve2}
\end{figure}

Using the same set of mass models as shown in figure \ref{f:mass_contours}, we can also compare our model solutions in terms of the values of the source distances that the model produces for each set of multiple images. In Figure \ref{f:f_k},the green points represents our model predicted $f_k$ curve assuming different redshift from 2 to 10.8 of JD. There each point represents the best $f_k$ using equation \ref{eq_f_k_equation} for all the images of each system, while the blue curve represents the theoretical $d_{ls}/d_s$ curve derived via the standard cosmological parameters. The derived values of $f_k$ are in very good agreement with their expected values given by the blue curve, except for system 3 that falls far from the line when system 6 is assigned a low redshift. The reason for this discrepancy is that the images of system 3 lie very close to system 6 such that the model cannot simultaneously accommodate both systems if they lie at low redshifts. The lower redshift photometric solution for system 6 generates a contradiction for the model by approximately 2$\sigma$, which is alleviated by setting system 6 to higher redshifts where it comes into very good agreement with the theoretical relation. Our best mass model assuming JD at $z=10.8$ is shown in Figure \ref{f:mass}, \ref{f:critcurve1}, and \ref{f:critcurve2}.

\subsection{Delensed Centroids Offsets}\label{sec:delensed_centroids}
\begin{figure*}
\centering
\includegraphics[width=150mm]{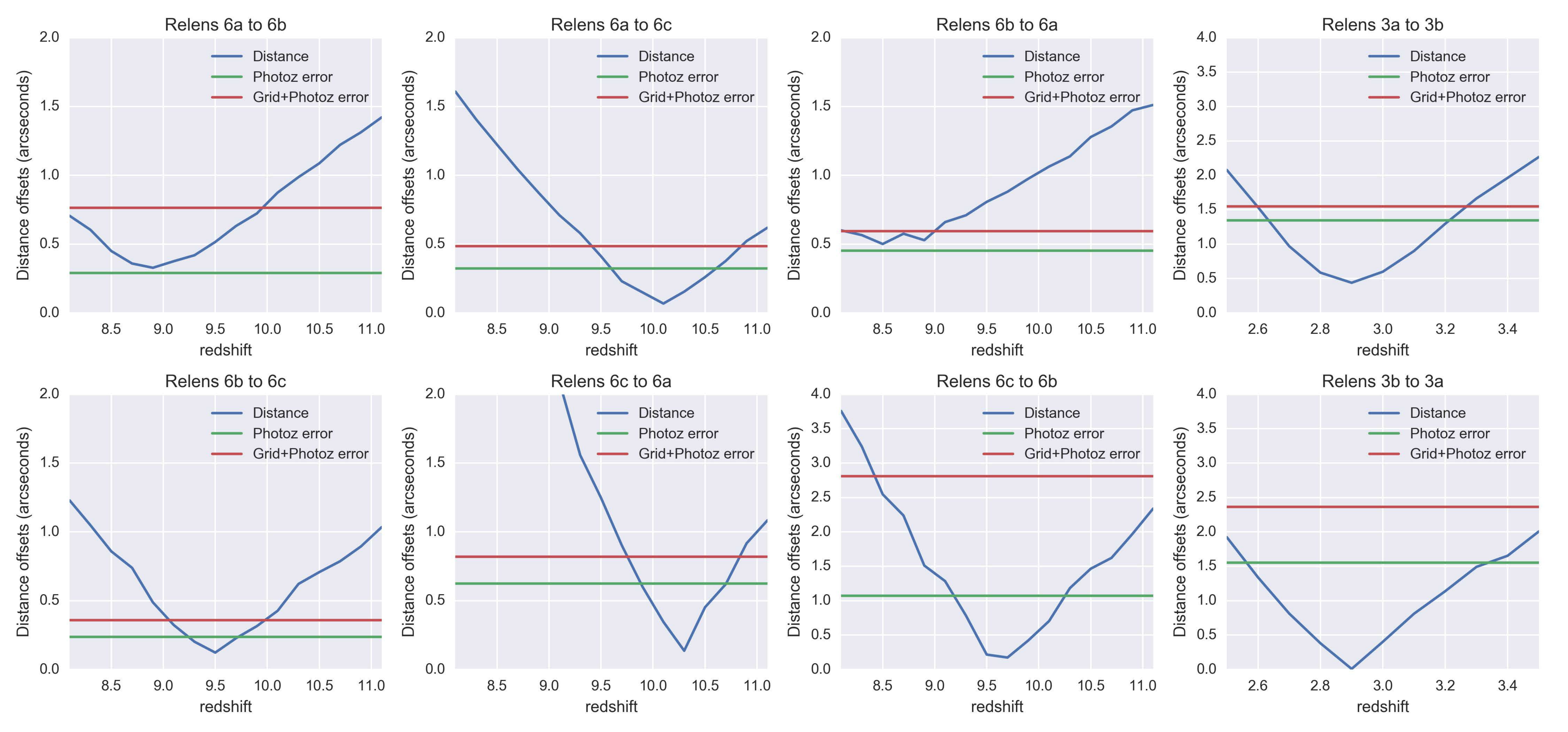}
\caption{
The relensed centroid offsets of system 6 and system 3 for a range of redshifts. The horizontal lines show the error estimated by using photometric redshift error alone and the photometric redshift error added to the grid error.
}
\label{f:geo-z-6-3}
\end{figure*}

\begin{figure}
\centering
\includegraphics[width=85mm]{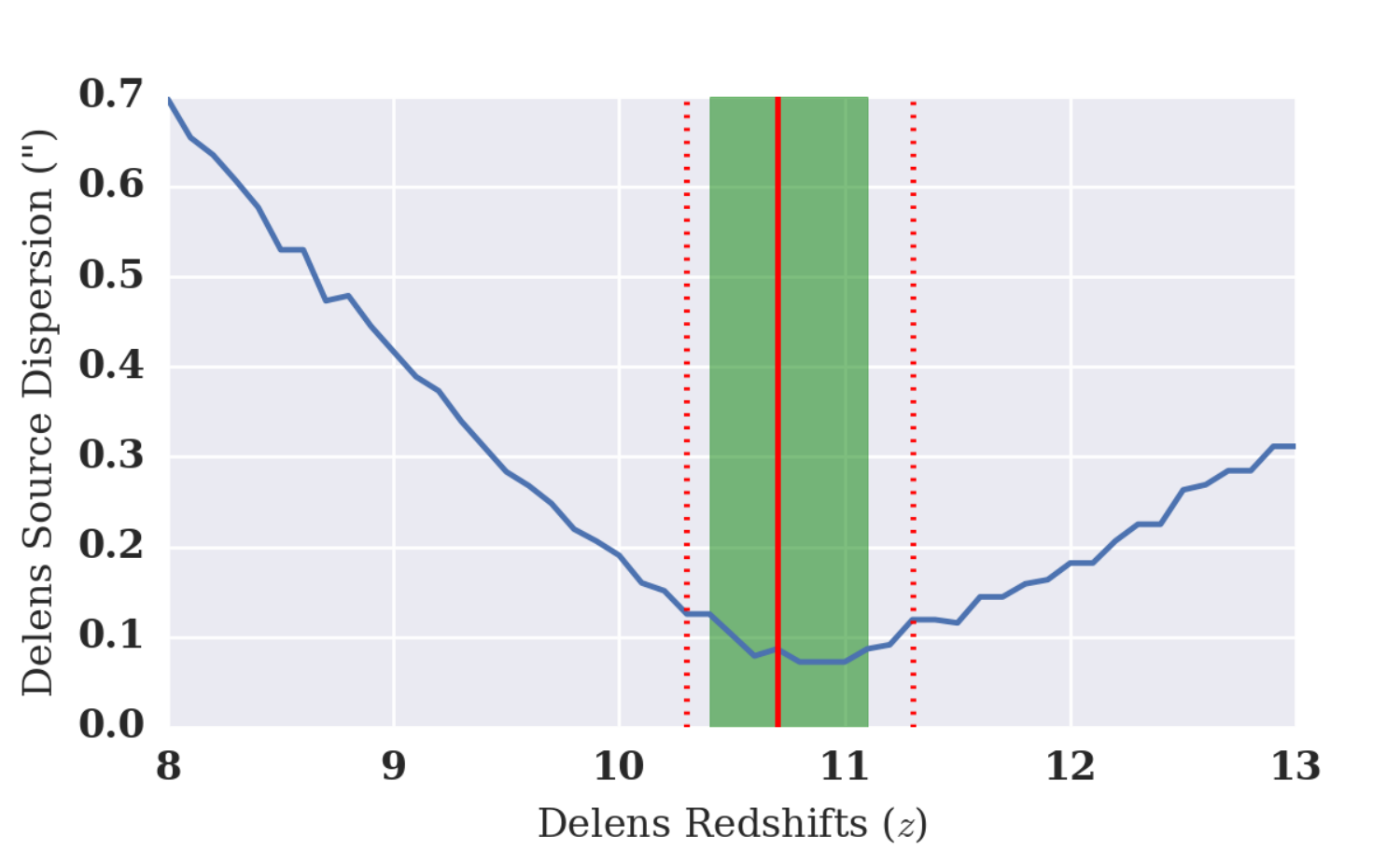}
\caption{
The delens centroid offsets versus delens redshift of system 6. The green region shows the range of minimum points we get for different models when we vary the photometric redshift, grid size and initial condition, which is very consistent with the photometrically derived redshift for system 6 in \citet{coe13}. The photometric redshift derived from \citet{coe13} was marked by the solid red line, while the upper and lower limits of the photometric redshift were marked in dotted red lines for comparison.
}
\label{f:delens_6}
\end{figure}

Here we describe the method we used to quantify the geometric redshift of system 6. In principle we could ``relens'' a lensed image of this source to the positions of its counter images. From which we could obtain a complete set of permutation of relens image $i$ to image $j$ where $i$ and $j$ are IDs of multiply lensed images of the same source (See Figure \ref{f:geo-z-6-3}). A similar, but more simple analysis could be done using the ``delens'' result. Using the deflection field calculated from our mass model, we ``delens'' all 3 images of MACS0647-JD to the source plane assuming the object has different redshifts. We quantify the dispersion of the delensed images as the average distance of individually delensed images to the barycenter position of all 3 delensed images. The result is shown in Figure \ref{f:delens_6}. Using this method, we find a formal redshift for JD of 10.8 with an rms uncertainty of +0.3 and -0.4 from Monte Carlo method.

\subsection{Comparison of Relensed Centroids Offsets}\label{sec:relensed_centroids}

\begin{figure}
\includegraphics[width=85mm]{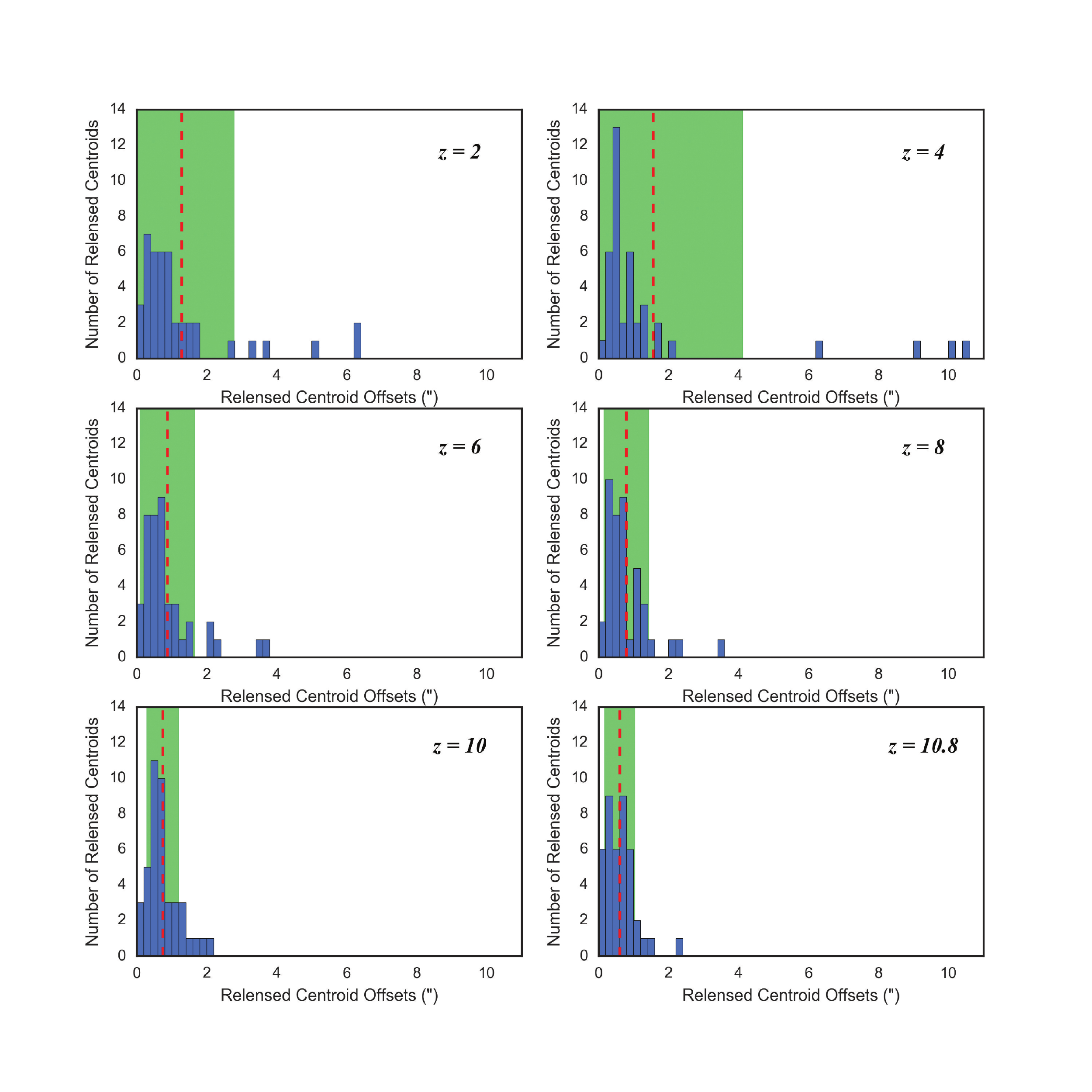}
\caption{
The relensed centroid offsets expressed in histograms, each presenting different mass model. In generating different mass models, the only parameter varied is the redshift of system 6.
}
\label{f:offsets}
\end{figure}

The self-consistency of our mass model is demonstrated by delensing and relensing individual images to predict the locations and fluxes of counter-images. Here, we focus on system 6 and the neighbouring system 3. The positional offsets between predicted and observed images are shown in Figure \ref{f:offsets}. Six individual mass models have been generated by our program, assuming $z=2,4,6,8,10,10.8$ respectively for system 6 and keeping all other systems redshifts fixed at the best photometric redshifts as shown in Table \ref{t:image_data}. As demonstrated by these histograms and Table \ref{tab:relens_stats}, the  self-consistency clearly increased when the assumed redshift for system 6 lies at higher redshift, resulting in a smaller mean difference between the observed and predicted image positions when averaged over all the multiply-lensed images. This tendency provides strong support for the high redshift of system 6.

\subsection{Individual multiply-lensed systems}\label{sec:individual}

\begin{figure*}
\centering
\includegraphics[width=150mm]{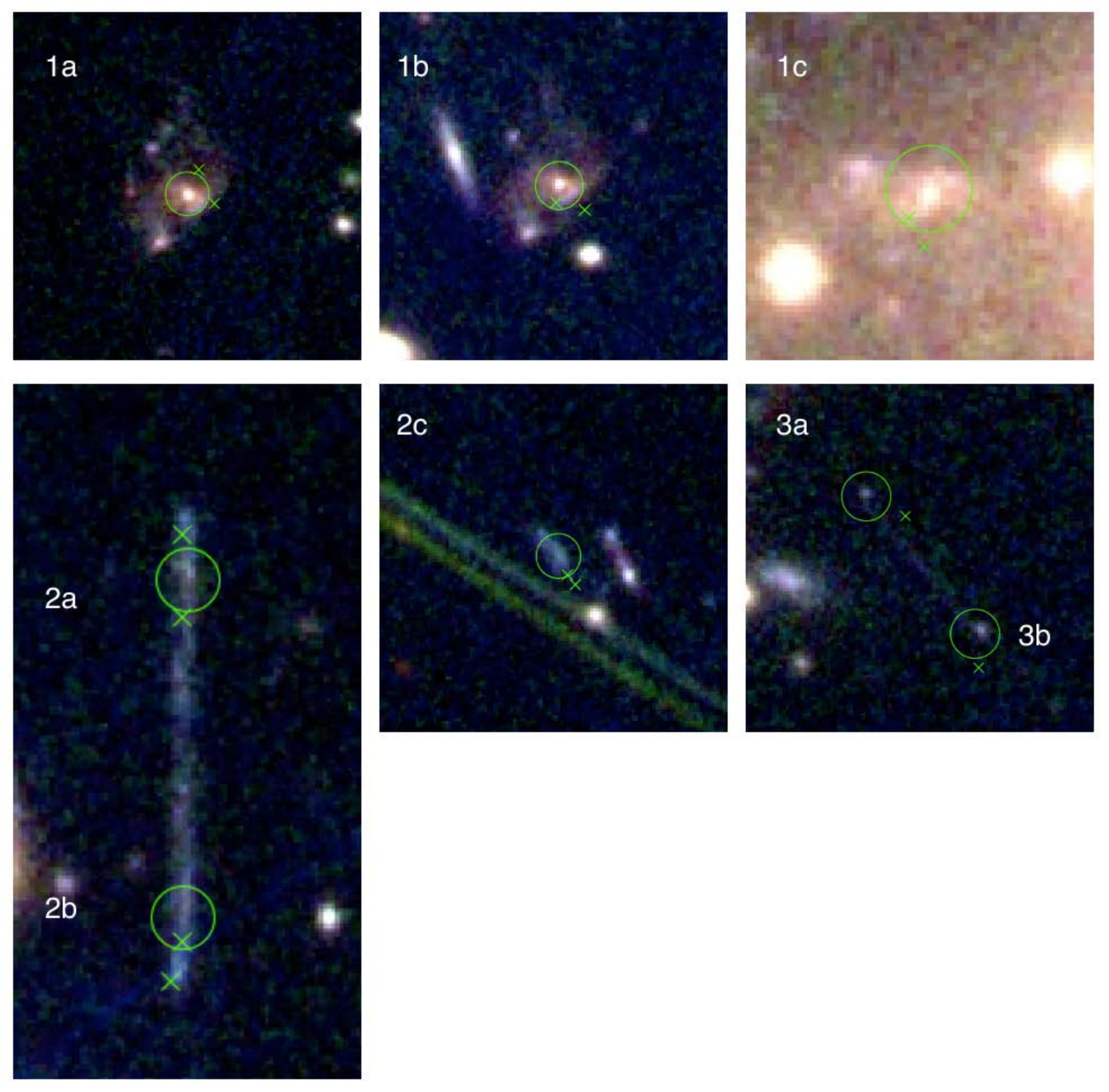}
\caption{
Image positions predicted by our lens model (x's) for systems 1 to 3. 
Each cross is obtained by delensing and relensing the centroid of the other counter images 
and with photometric redshift used as input. 
The circle is centred at the observed image centroid position, with a radius of 0.5''.
}
\label{f:stamps1}
\end{figure*}

\begin{figure*}
\centering
\includegraphics[width=150mm]{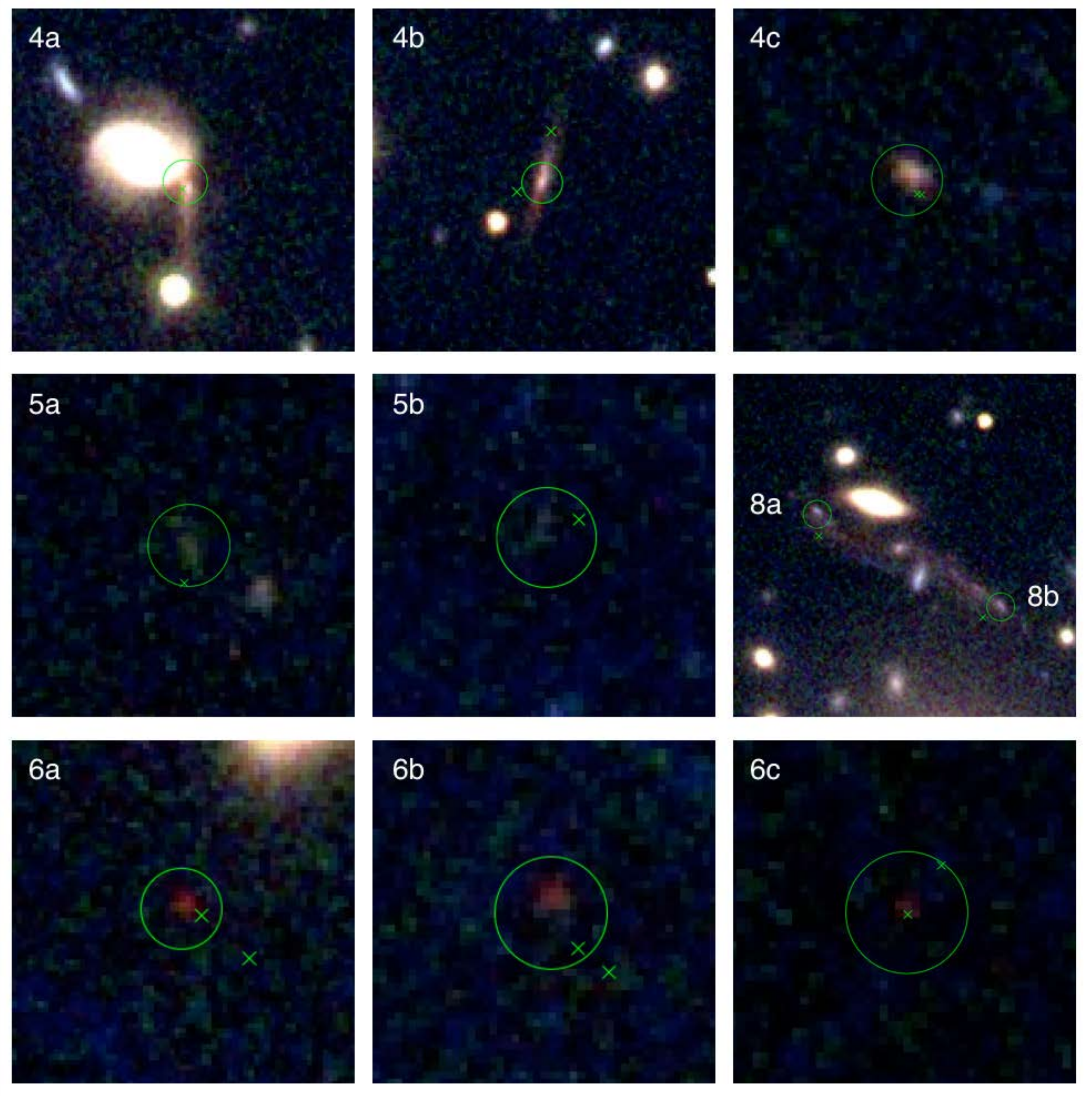}
\caption{
Same as figure \ref{f:stamps1} except here we show systems 4, 5, 6 and 8.
}
\label{f:stamps2}
\end{figure*}

\begin{figure*}
\centering
\includegraphics[width=150mm]{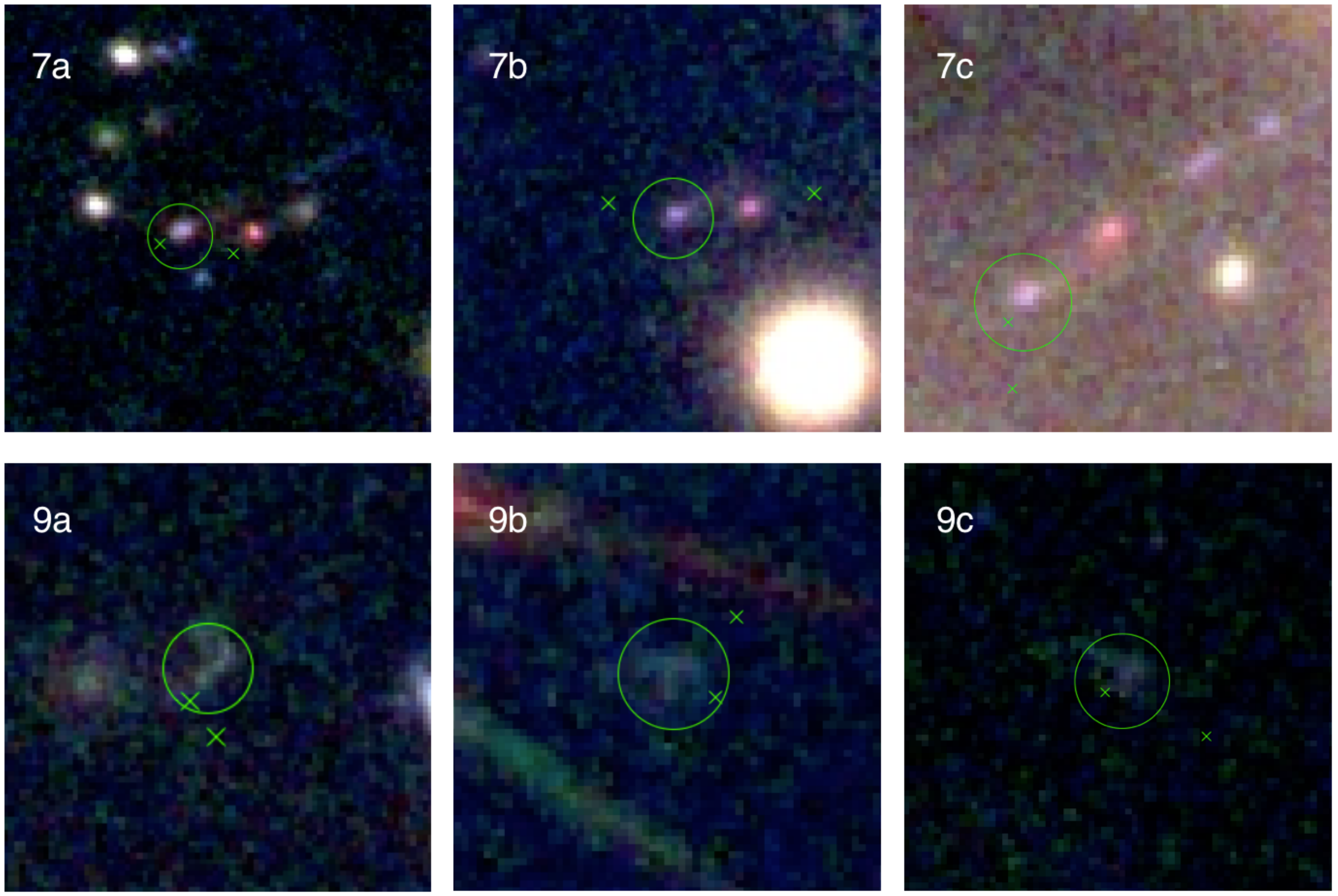}
\caption{
Same as figure \ref{f:stamps1} except here we show systems 7 and 9.
}
\label{f:stamps3}
\end{figure*}

\begin{figure}
\centering
\includegraphics[width=85mm]{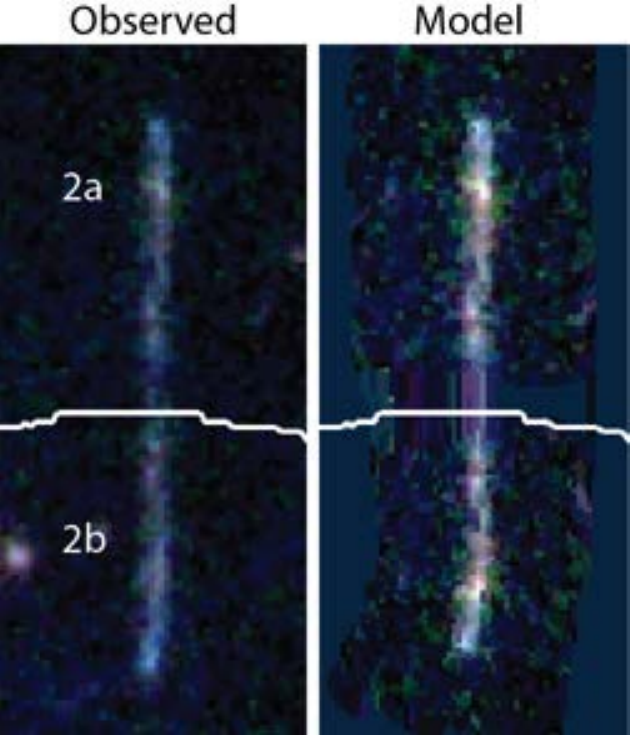}
\caption{
Here we provide our own model reconstruction of image 2a and 2b from system 2, using the pixels of image 2a to generate the other image for comparison with the observed images. This reconstruction is made by solving for the lens equation, and then delensing followed by relensing of the observed flux in each pixel. Notice for image 2b, our mass model is even able to reproduce the small left curve of the observed image. The white curve shows the critical curve at the redshift of system 2.
}
\label{f:system_2}
\end{figure}

Here, we describe the multiply-lensed systems used to derive the lens model for MACS0647. The ID number, positions, redshifts are tabulated in Table \ref{t:image_data}. The magnification predicted by our lens model and photometry of the individual lensed images are tabulated in table \ref{t:luminosity}. The accuracy of our identifications is demonstrated by the relens results of the well resolved images that have notable elongation/distortion and distinct internal structures, as well as other multiply-lensed images that are not well-resolved. We also calculate the centroids of each predicted counter images in figures \ref{f:stamps1} to \ref{f:stamps3}. 

System 1.  
This triply lensed spiral galaxy is the best resolved system. The images are bright and well-resolved and we obtain very accurate positions and redshifts for this  galaxy as expected with model relensed centroids that are all in good agreement with the data.

System 2.  
This is an image system with 2a and 2b very close to the critical curve, where the two images almost meet to form a long vertical line. Image 2c is rather isolated and appears almost unresolved.
As seen in figure \ref{f:system_2}, delens and relens 2a and 2b successfully reproduced the long vertical line very similar to observation. However, relens 2c produced a curved line in the region of 2a and 2b.

System 3. 
This system lies very close to MACS0647-JD, the high redshift candidate. There are two identified images, quite close to each other. Our mass model predicts a third image at the top left part of the field of view, as labeled in Figure \ref{f:combo_map}.  After searching through the data, we could not find the image without ambiguity. Our predicted brightness for this image is comparable with the noise in 
the F475 band where it should be strongest (AB magnitude $\sim$ 29.4) that is consistent with its lack of detection. We do not include any third image candidate into our constraints.  As the system is so close to our $z\sim11$ candidate, it serves as a good indicator of whether $z\sim11$ is a reliable redshift of system 6 using the ``geometric'' method, as we demonstrated above.

System 4. 
This system is a triply lensed galaxy. One of the multiply lensed images, 4a, lies closely to a cluster member galaxy. Our model successfully delens and relens other images to the position of 4a. Further investigation on the morphology of that image can be done to constrain the mass model of the cluster member galaxy.

System 5. 
A very faint doubly lensed image system, which is the second highest redshift galaxy identified in the lensing field. 

System 6.  
The second highest redshift galaxy ever observed, and the highest redshift lensed galaxy identified. Using our best mass model, we predict a geometric redshift of $z\sim10.8$, coincide with the photometric redshift derived by \citet{coe13}.

System 7. 
This is a triply lensed spiral galaxy lies near to system 1. The images are well resolved, and slightly fainter than system 1. The source has the same photometric redshift as system 1 but with slight offset in position. 

System 8. 
This doubly lensed system exhibits major lensing effect from a nearby cluster member galaxy. As shown in Figure \ref{f:stamps2}, our model is able to reproduce the relensed centroids of both multiple images. The relens result predicts a third image 8c at a position near image 1b and 7b, with predicted brightness similar to that of 8a. With careful inspection at the predicted region, we could not find the image without ambiguity. As 8a and 8b lie closely to the critical curve, the morphology is highly distorted. For this reason, it is more challenging to identify the counterimage.

System 9.  
This system is amongst the faintest galaxies in the lensing field. With hugely separated positions of multiply lensed images, it provides useful constraints to the mass model.

\section{Lens Model Prediction Using Relative Brightnesses.}\label{sec:mag_prediction}

Having demonstrated above the reliability of the high redshift option for MACS0647-JD, we now make use of 
its redshift to evaluate the predictive power of our lensing model by looking at the relative fluxes of all 
the multiply lensed systems. The evaluation was first made in \citet{broadhurst05} for the case of A1689 and 
recently for the first HFF cluster A2744 by \citet{lam14} using our free-form WSLAP+ code. Note that the information contained in the relative brightness of individual sets of lensed images is not used when constructing  the lens model.
 
\begin{figure}
\centering
\includegraphics[width=85mm]{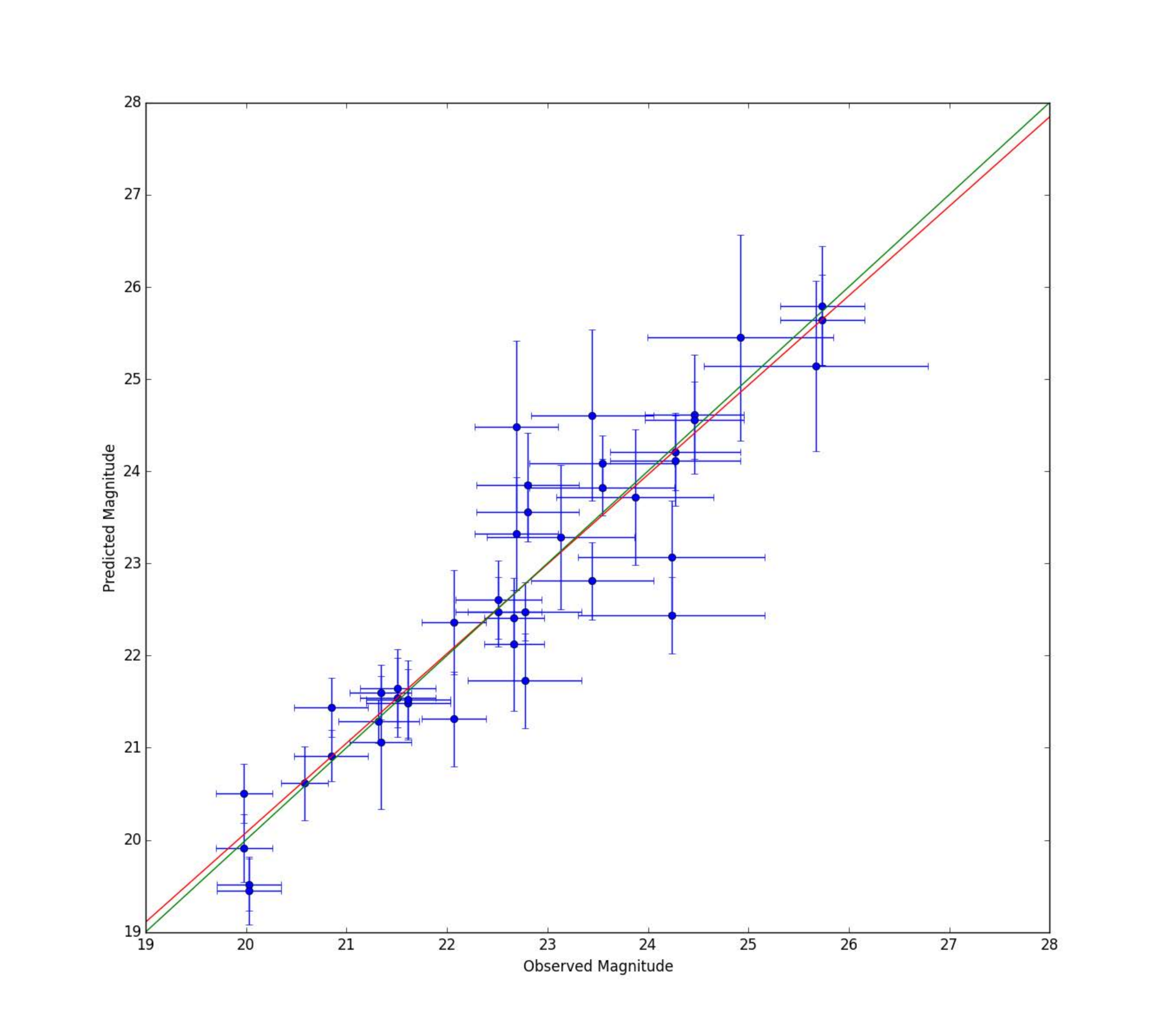}
\caption{
Relative magnitudes predicted by our model plotted against observed magnitudes of multiple images. 
The predicted magnitudes are calculated by magnifying the observed magnitudes of the first images in each system by our lens model. 
}
\label{f:mag1}
\end{figure}

\begin{figure}
\centering
\includegraphics[width=85mm]{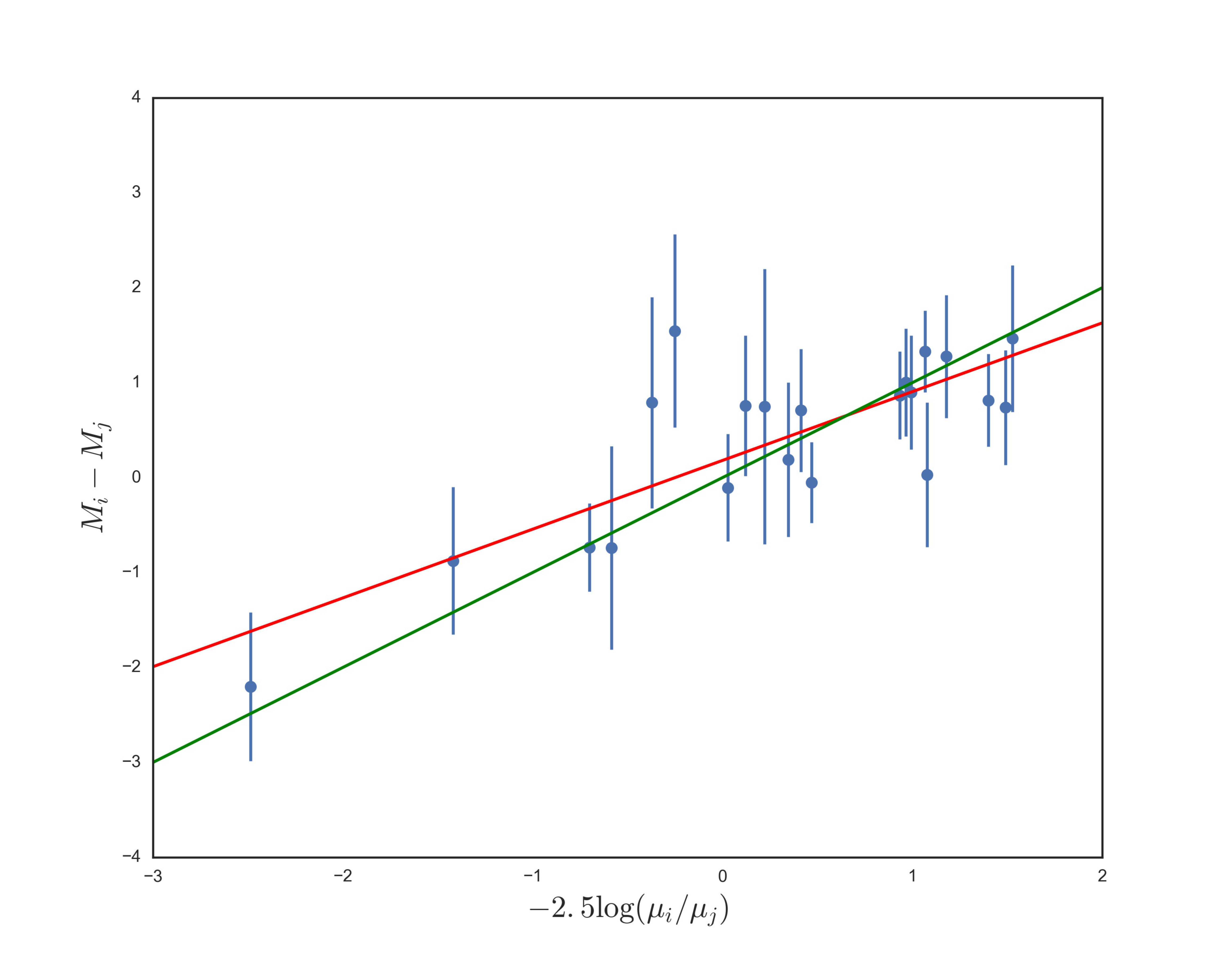}
\caption{
The relative magnitude difference between two images $i$ and $j$ of the same source versus $-2.5\log(\mu_i/\mu_j)$. The green line shows the 1-to-1 correspondence line and the red line shows the best-fit line from our data.}
\label{f:mag2}
\end{figure}

In figure \ref{f:mag1} we plot the model-predicted relative magnitudes against the observed magnitudes for the images that constrained the reconstructed lens model.  The magnitudes are predicted by magnifying or de-magnifying the observed magnitudes of the first and second images in each system, as 
typically there are three images per system with usually one case where the photometry is poor due to overlap with a member galaxy.  Reassuringly, we find a clear linear relation between the predicted and observed magnitudes, indicating the high level of predictive power of our model. There are a few outliers attributable to the proximity to the critical curve to the ended image so that the correction is relatively large,  but in general the overall scatter is small and the trend shows no systematic deviation from linearity.  The scatter in this relation, however, significantly exceeds the photometric uncertainty. We are currently examining how we can incorporate the relative brightness of the lens images as additional constraints when modeling the cluster lens. We present these results in (Figure \ref{f:mag2}), where the observational parameters are compared to the model dependent parameters and we provide our predictions for the intrinsic brightness of MACS0647-JD in table \ref{t:delens_6}.

\section{Geometric Redshift Consistency}\label{sec:geoz_consistency}

\begin{figure}
\centering
\includegraphics[width=85mm]{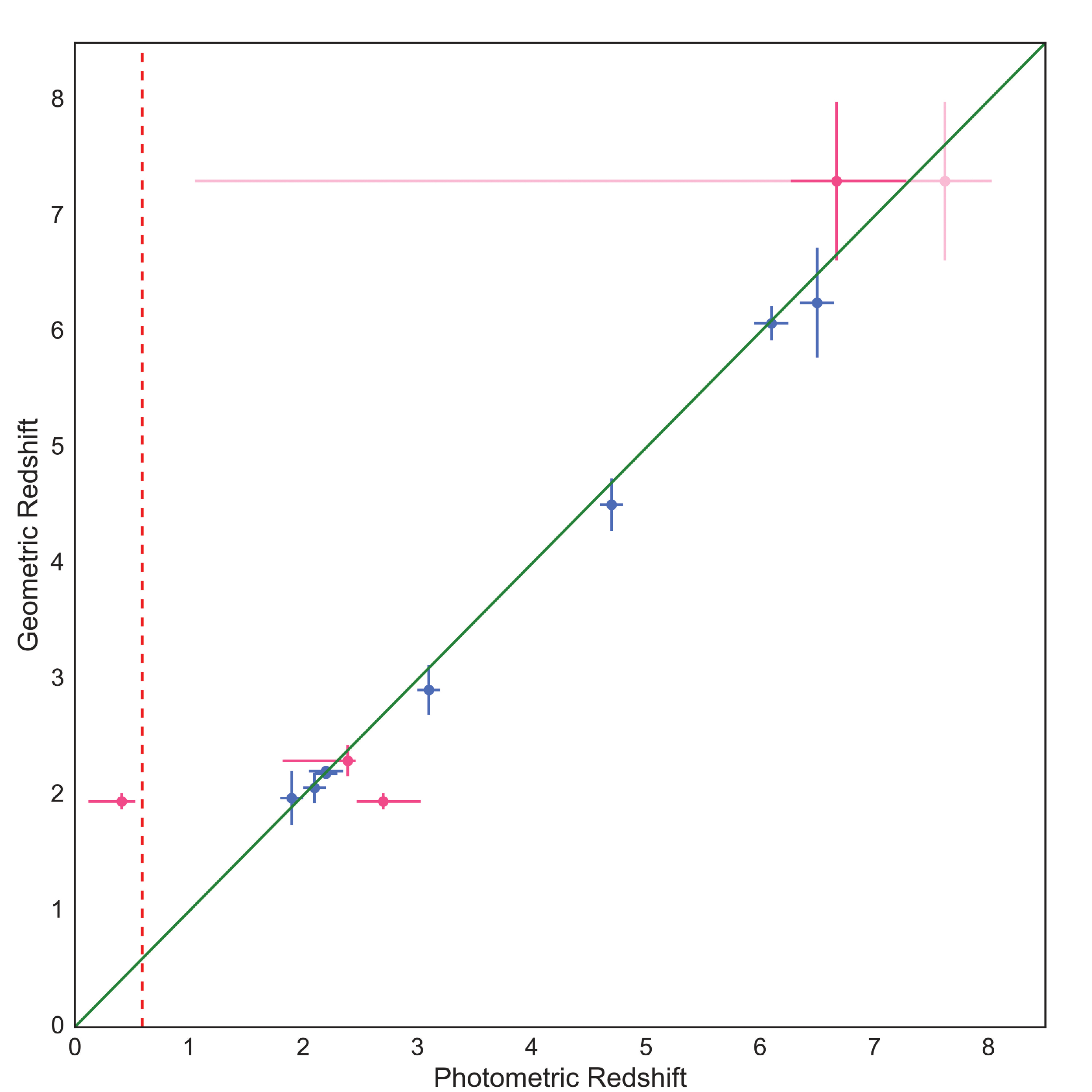}
\caption{Here we show the predicted geometric redshifts and the photometric redshifts of all the multiply-lensed systems . Values and errors of photometric redshifts obtained from \citet{coe13} are shown in blue points. Three systems identified by \citet{zitrin15a} are shown as pink points, with each point representing an individual counter image. The faintest pink point shows the result for image 10.2, which has a relatively large error on its photometric redshift measurement. The green line is the one to one correspondence line, while the dash line shows the redshift of the lensing cluster. The values and errors in geometric redshift is calculated by averaging 9 reconstructed mass models for testing each lensed system, where with the system in question is not included in the construction of the set of mass model with which its geometric redshift is evaluated. The red dash line shows the redshift of the lensing cluster MACS0647. From the graph, the image with photometric redshift $z=0.41$ cannot be lensed as it is in front of the lensing cluster. In such case geometric redshift is useful to estimate the redshift of a galaxy.
}
\label{f:photoz_vs_geoz}
\end{figure}

Here we look in turn at each system of multiple images and ask how well the geometric redshift that we derive for it agrees with the measured photometric redshift. This can be done by constructing a set of lens models where we leave out each system in turn and then use that corresponding model to estimate the geometric redshift of that system that was left out of its construction. The result of this procedure is shown in figure \ref{f:photoz_vs_geoz}, where we plot the predicted geometric redshift of all systems against their photometric redshifts as measured by \citet{coe13}. To compute their geometric redshifts, we reconstructed 9 different mass models, each with one of the systems not included in the initial constraint. We then compute the average and standard deviation of the predicted geometric redshifts out of all 9 models. As it can be seen from the figure, the data points lie close to the correspondence line, with errors within one sigma. The consistency between the photometric and geometric redshifts is striking and indicates that there is no obvious outlier with an unreliable photometric redshift. 

\section{Other Candidate Lensed Systems}\label{sec:other_images}
Apart from 9 secured lensed systems as identified in \citet{coe13}, three more candidate lensed systems were identified by \citet{zitrin15a}. The details of these candidate systems are shown in table \ref{table:zitrin_redshifts}

System 10.
This system consists of two multiply lensed images. A slight difference is observed for the most likely photometric redshift of counter images. The redshift error of image 10.2 is quite large related to ambiguity in the SED fitting. Our geometric redshift agrees well with the photometric redshift of image 10.1.

System 11.
There is a huge disagreement in photometric redshifts between the counter images. The geometric redshift predicted for this system lies at $z=1.95$, corresponding approximately to the average of the photometric redshift measured. 

System 12.
Although photometric redshfit is not available for 12.1, there is a good agreement between 12.2 and 12.3. The geometric redshift predicted agrees very well with the photometric redshift.

\section{Discussion}\label{sec:discussion}

\begin{figure}
\centering
\includegraphics[width=85mm]{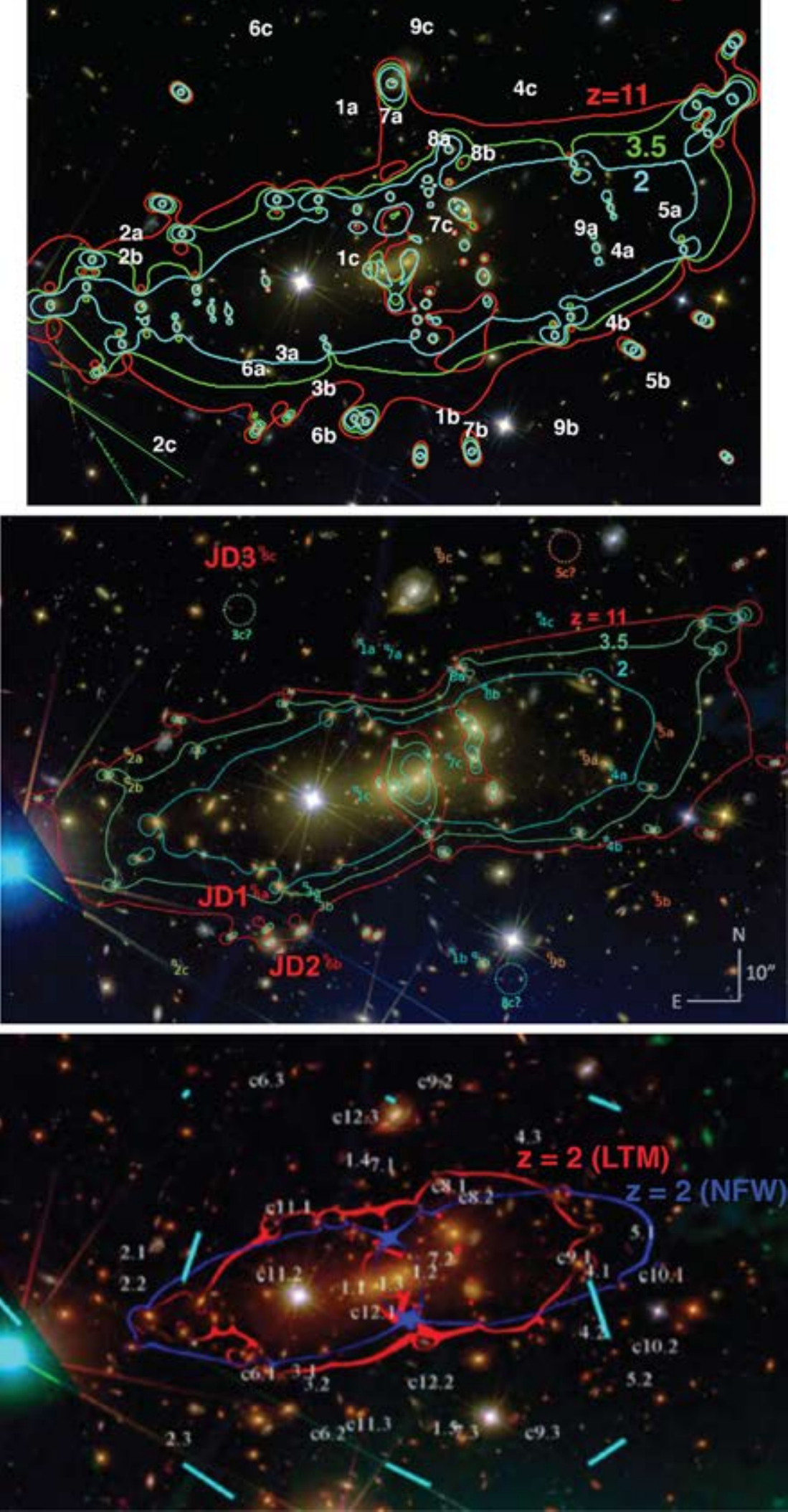}
\caption{Here we compare our critical curves for our best fit free-form model (top) with those from \citet{coe13} (middle) that uses the parametric Lenstool model, and also a pair of solutions taken from \citet{zitrin15a} for which two good solutions were found (bottom). Note that in all cases the size of the critical curves increases with redshift and the corresponding redshifts are marked. Note that the more elliptical model of \citet{zitrin15a} agrees well with the shape of our critical curves and those of \cite{coe13}.
}
\label{f:critcurve_comparison}
\end{figure}

Previous lens modelling has not provided a definitive geometric redshift for system 6, or indeed for any other system lensed by this cluster. In \citet{coe13}, the larger critical curve at $z=11$ does pass centrally between images 6a and 6b like in our model and the critical curves are fairly similar (see Figure \ref{f:critcurve_comparison}) so that we can imagine the form of parametric modelling might lead to consistency in this respect. In \citet{zitrin15a} an attempt was made to find good solutions in the context of the ``light traces mass'' approach, which has a degree of flexibility and a minimum of free parameters. Two solutions were presented by \citet{zitrin15a} as shown in Figure \ref{f:critcurve_comparison} where the more elliptical of the two cases favours high redshift. Our free form model has the desirable approach of providing a definitive geometric redshift for all the lensed galaxies here that are not dependent on profile parameters but to a free forms mass distribution that is very general. Note also, as discussed in section \ref{sec:delensed_centroids} and \ref{sec:geoz_consistency},  the geometric redshifts we determine for all systems in this work are in good agreement with their independently determined photometric redshifts. Furthermore, the predictive power in getting the relative magnifications of lensed images were demonstrated in section \ref{sec:mag_prediction}.

In addition to the high redshift galaxy that we have analysed here
finding a high geometric redshift of $z\simeq 10.8^{+0.3}_{-0.4}$ for MACS0647-JD,
the same free form method has been successful in determining the 
geometric redshifts of Hubble frontier field galaxies, including a triply lensed galaxy 
with a photometric redshift of  $z\simeq 9.6$ behind A2744 and 
a confirming geometric redshift \citep{zitrin14} from our free form method. 

MACS0647-JD is the highest redshift galaxy so far detected from lensing despite 2 magnitudes of
magnification and the great depth of this imaging, no lensed galaxy
is yet discovered beyond the most distant galaxies already
known for several years \citep{coe13,coe15,ishigaki15,oesch13,zitrin14,zheng12}. This difficulty is not due to the filter
choice, which in principle can access Lyman-break galaxies
out to $z\simeq 12$. As the HFF program progresses to completion
we may anticipate a clear, field averaged constraint on
the number density of galaxies lying above $z > 9.0$. A significant
absence of such galaxies is not predicted for the LCDM
model, where several galaxies are expected per HFF cluster
in the range $9 < z < 10$ on the basis of the standard LCDM
model, by extrapolating the luminosity function (e.g \citet{coe15,schive16}).

The most recent work by \citet{oesch15}, \citet{bouwens16},  in relation to the deep field surveys has 
indicated a surprisingly complex picture that confirms the  rapid decline in
the integrated star formation  above $z>8$, exceeding the extrapolated rate from the
well defined $3<z<8$ range together with several new claims of relatively relatively bright field galaxies at $z>10$ including one with $z\simeq 11 .1$ supported by HST grism observations. A clear dearth of lower luminosity galaxies may also indicate in the latest \citep{bouwens15} work at $8< z <10$ and this is in contrast to the rising luminosity functions reported by \citet{dunlop16},  for the HFF and other deep fields, though this rests on the reliability of very faint sources near the flux limit. It is interesting to speculate that the
enhanced gas cooling is indicated in the case of wave-DM \citep{schive14} due to the massive dense solitonic core that form first in this context, quite unlike that of LCDM where the smallest galaxies form first with very inefficient star formation expected as winds readily remove hot gas 
from these small first halos \citep{franx97,frye98,pettini98,frye00,ferrara13,scannapieco00}. In this axion-like interpretation of CDM structure forms in the same way as standard heavy particle CDM on large scales \citep{schive14} but is suppressed below an inherent Jeans scale in the dark matter \citep{sikivie09} so that the onset of galaxy formation is delayed with respect to CDM \citep{bozek14,schive14, schive16}. Whereas the onset of AGN activity may be relatively earlier than LCDM, because within $\psi$DM halos, a dense ``solitonic'' standing-wave core forms initially, of up to  $10^{10}M_\odot$, capable of focussing gas, seeding and feeding early super massive black holes \citep{schive14}. 

In addition, expectations regarding galaxy formation are also in flux because of the latest downward revision in the redshift estimate for reionization at $z\simeq 8$ from Planck polarisation measurements of the optical depth of electron scattering of the CMB \citep{planck16}. Empirically, this low redshift squares with the contested claim of an accelerated decline in the galaxy luminosity density at $z>8$ \citep{bouwens15,mcclure16,oesch13}. These deepest observations with the Hubble and Planck satellites may together be consistent in indicating a later and more sudden onset of galaxy formation than widely anticipated. Empirically, it appears that consistency is seen between the downwardly revised ``instantaneous redshift" of reionization,  $z\simeq 8$,  when most of the cosmologically distributed hydrogen had become reionized \citep{planck16}  with the tentative claim of an accelerated decline in the integrated star formation rate at $z>8$ \citep{oesch10, oesch13} and supported by a absence of any galaxy detected beyond $\simeq 11.0$ \citep{oesch16}, despite increasingly deep Hubble Frontier Field imaging \citep{diego15a,diego15b,diego16b,diego16a,lam14,zheng12}. Late reionization is also supported by a marked decline above $z \gtrsim 7$ in the incidence of Ly-$\alpha$ emission from high redshift galaxy spectroscopy,  that can be reproduced by photoionizaton models with a much higher neutral fraction than anticipated \citep{choudhury15,mesinger15} indicating later galaxy and QSO formation \citep{madau15}.

At face value all these observations now seem to be at odds with LCDM for which no preferred redshift of galaxy formation is anticipated as the growth of structure is scale free for collisionless, cold dark matter, so that a smooth decline in galaxy numbers is widely anticipated to beyond $z>20$ with the first stars predicted formed at $z\simeq 50$ \citep{naoz06}. For the same reason, the low optical depth emerging from Plank satellite CMB work also seems in tension with CDM.  Earlier higher estimates of $\tau$ were rather too large to be comfortably provided by galaxy photoionisation alone within the context of LCDM with ``reasonable'' photoioization parameters, including the unknown UV-escape fraction and the hardness of ionising radiation spectrum \citep{robertson12,robertson14}.  However, a large variation in forest transparency between the highest redshift QSO sightlines \citep{becker15} may point to relatively few UV sources and hence to AGN rather than star formation  as the dominant  source of photoionisation \citep{madau15}, with the necessary ionisation may be provided by low luminosity AGNs forming a flatter redshift evolution for low luminosity QSOs in the $4<z<6$ range \citep{glikman11,giallongo15} as pointed out by \citep{madau15}. In this case one may either conclude
that the escape fraction from early galaxies is negligible, or that such galaxies are not formed and this possible absence is now supported by the lack of low luminosity galaxies at $z>8$,
with important implications for the the nature of dark matter, as described above.
 
\section{Summary, Conclusions and Future Work}\label{sec:conclusions}

We have focussed here on the geometric constraint that lensing can provide for the distance of
the highest redshift galaxy claimed in deep lensing work. We show quite generally with our free-form lensing model that its geometric redshift must lie well above the lower redshift neighbouring system at $z\simeq 3.0$ (system 3). Spectral confirmation has proven infeasible to date, but alternative solutions such as $z \sim 2.5$ dusty galaxies have been  disfavored with high confidence \citep{coe13} Furthermore we find that our free form lens model based on all 9 sets of multiple images clearly prefers a high redshift for MACS0647-JD, on grounds both of its position and the relative image brightnesses. Our work clearly supports the high redshift interpretation, with an independently determined geometric redshift that we have presented here of $z\simeq 10.8^{+0.3}_{-0.4}$. We also obtain an intrinsic luminosity of $M_{UV} = -19.4$ (corresponds to $L / L^\ast_{z=3} = 0.22$) that is close to that of \citet{coe13} but without associated model dependance. The calculated luminosity is a factor of a few lower compared with the objects mentioned in \cite{holwerda15} Our work also support two of the three  new multiply lensed candidates tentatively identified by \citet{zitrin15a}, which we find show  agreement between their geometric and photometric redshifts.

We have uncovered a satisfyingly tight agreement between our photometric redshifts and the ``geometric redshift'' that we derive for each multiply-lensed system, as described in section \ref{sec:geoz_consistency} and shown in table \ref{t:image_data}. This good agreement indicates a high degree of self-consistency in the model, going well beyond the usual scatter in this type of figure, \citet{broadhurst05, halkola06} for A1689 and explored also in a more model dependent way by \citet{limousin07} for the NFW profile. Hence, the free-form model-independence of our method can in principle permit a joint constraint on the mass distribution and cosmology. The feasibility of this  has been examined with simulations by \citet{lubini14}, and now seems warranted by our improved observational  precision achieved here. We have also examined the self consistency of the model by comparing the  predicted and observed brightnesses of the lensed images as shown in Figure \ref{f:mag1}. In principle we could also include this magnification related information when constraining the model, which will be done in future analysis.

This is the second such attempt to use the brightness data in a model independent way, like the earlier work of \citet{lam14}  which utilised the same WSLAP+ code. We find a tight linear relation between the predicted relative magnitudes and the relative observed fluxes, demonstrating the predictive power of our lens model. This has not generally been found using parameterised model where the form of the radial profile of the cluster is limited to a narrow family of models \citep{caminha16b} and further underscores the utility of our free form approach in the case where the abundance of deep data justifies this method, over that of parameterised models. Parameterised models can only be partially appropriate at best for cluster mass distributions, particularly in the case of merging clusters where the complexities of tidal effects during encounters means the general mass distribution cannot be expected to adhere to a sum of idealised elliptical, power-law mass halos usually adopted.  The improvements we have found here in using relative fluxes and local galaxy separations to constrain geometric distances motivates further exploration of the best means of combining the independent positional and flux data, for breaking the degeneracies allowed by positional data alone \citep{schneider14}
with  the goal of examining the cosmological-distance redshift relation in the unexplored range $z>2$, beyond the SNIa work, with our model-independent lensing tools.

\acknowledgements
$\it{Acknowledgements.}$
The work is based on observations made with the NASA/ESA {\it Hubble Space Telescope} and operated by the Association of Universities for Research in Astronomy, Inc. under NASA contract NAS 5-26555. 
T. J. Broadhurst gratefully acknowledges the Visiting Research Professor Scheme at the University of Hong Kong.
J. M. D. acknowledges support of the consolider project AYA2015-64508-P (MINECO/FEDER, UE), CAD2010-00064 and AYA2012-39475-C02-01 funded by the Ministerio de Economia y Competitividad, and support from the Department of Physics at The University of Hong Kong for a visit to work on this project.
J. L. acknowledges support from the University of Hong Kong via a Seed Funding Program for Basic Research through grant 201411159166 at the start of this project, thus enabling follow-up support from the Research Grants Council of Hong Kong through grant 17319316.

\clearpage
\bibliography{citation}

\clearpage


\begin{deluxetable*}{cccccp{4cm}}
\centering
\tabletypesize{\scriptsize}
\tablecaption{Detailed information of individual lensed images\label{t:image_data}}
\tablewidth{0pt}
\tablehead{\colhead{Image} & \colhead{RA} & \colhead{Dec} & \colhead{BPZ (image used)} & \colhead{Geo-z} & \colhead{Remarks}}
\startdata
1a & 06 47 51.87 & +70 15 20.9 & 2.2$\pm$0.1& 2.2 &  \\
1b & 06 47 48.54 & +70 14 23.9 & 2.2$\pm$0.1& 2.2 &  \\
1c & 06 47 52.01 & +70 14 53.8 & 2.2$\pm$0.1& 2.2 &  \\
2a & 06 48 00.33 & +70 15 00.7 & 4.7$\pm$0.1 & 4.4 & \\
2b & 06 48 00.33 & +70 14 55.4 & 4.7$\pm$0.1 & 4.4 & \\
2c & 06 47 58.62 & +70 14 21.8 & 4.7$\pm$0.1 & 4.4 & \\
3a & 06 47 53.85 & +70 14 36.2 & 3.1$\pm$0.1 & 3.1 & \\
3b & 06 47 53.41 & +70 14 33.5 & 3.1$\pm$0.1 & 3.1 & \\
4a & 06 47 42.75 & +70 14 57.7 & 1.9$\pm$0.1 & 1.9 & \\
4b & 06 47 42.93 & +70 14 44.5 & 1.9$\pm$0.1 & 1.9 & \\
4c & 06 47 45.37 & +70 15 25.8 & 1.9$\pm$0.1 & 1.9 & \\
5a & 06 47 41.04 & +70 15 05.5 & 6.5$\pm$0.15 & 6.5 & \\
5b & 06 47 41.16 & +70 14 34.4 & 6.5$\pm$0.15 & 6.5 & \\
6a & 06 47 55.74 & +70 14 35.7 & $10.7^{+0.6}_{-0.4}$ & 10.8 & Also known as JD1\\
6b & 06 47 53.11 & +70 14 22.8 & $10.7^{+0.6}_{-0.4}$ & 10.8 & Also known as JD2\\
6c & 06 47 55.45 & +70 15 38.0 & $10.7^{+0.6}_{-0.4}$ & 10.8 & Also known as JD3\\
7a & 06 47 50.91 & +70 15 19.9 & 2.2$\pm$0.15 & 2.2 & \\
7b & 06 47 47.73 & +70 14 23.2 & 2.2$\pm$0.15 & 2.2 & \\
7c & 06 47 48.69 & +70 14 59.8 & 2.2$\pm$0.15 & 2.2 & \\
8a & 06 47 48.61 & +70 15 15.8 & 2.3$\pm$0.1 & 2.2 & \\
8b & 06 47 47.34 & +70 15 12.5 & 2.3$\pm$0.1 & 2.2 & \\
9a & 06 47 43.79 & +70 15 00.4 & 5.9$\pm$0.15 & 6.2 & \\
9b & 06 47 44.98 & +70 14 23.2 & 5.9$\pm$0.15 & 6.2 & \\
9c & 06 47 49.06 & +70 15 37.7 & 5.9$\pm$0.15 & 6.2 &
\enddata
\end{deluxetable*}

\begin{deluxetable*}{ccc}
\centering
\tabletypesize{\scriptsize}
\tablecolumns{3}
\tablecaption{Re-lens Statistics of Models Assuming Different System 6 Redshifts\label{tab:relens_stats}}
\tablehead{\colhead{Assumed Redshift of MACS0647-JD} & \colhead{Mean Relens Offsets (")} & \colhead{SD of Re-lens Offsets (")}}
\startdata
2 & 1.28 & 1.50 \\
4 & 1.56 & 2.56 \\
6 & 0.87 & 0.79 \\
8 & 0.79 & 0.65 \\
10 & 0.73 & 0.45 \\
10.8 & 0.61 & 0.44
\enddata

\tablecomments{The numbers are also visualised as red dotted lines and green regions of Figure \ref{f:offsets}}
\end{deluxetable*}

\newpage
\begin{deluxetable*}{ccccccccp{4cm}}
\centering
\tabletypesize{\scriptsize}
\tablecaption{Magnification and photometry of individual lensed images\label{t:luminosity}}
\tablewidth{0pt}
\tablehead{\colhead{Image} & \colhead{$\mu$} & \colhead{F160W AB mag}}
\startdata
1a & 6.1 & 22.9$\pm$0.3\\
1b & 4.0 & 22.3$\pm$0.3\\
1c & 1.7 & 23.0$\pm$0.4&  \\
2a & 13.0 & 24.15$\pm$0.4\\
2b & 13.3 & 24.0$\pm$0.4\\
2c & 5.3 & 25.0$\pm$0.4\\
3a & 24.7 & 27.0$\pm$0.8 \\
3b & 43.0 & 26.2$\pm$0.7 \\
4a & 0.9 & 24.9$\pm$0.7 \\
4b & 8.6 & 23.6$\pm$0.3 \\
4c & 3.2 & 24.9$\pm$0.3 \\
5a & 3.6 & 27.6$\pm$0.9 \\
5b & 2.9 & 28.5$\pm$1.1 \\
6a & 7.9 & 25.9$\pm$0.1 \\
6b & 7.7 & 26.1$\pm$0.1 \\
6c & 2.6 & 27.3$\pm$0.1 \\
7a & 9.5 & 24.2$\pm$0.3 \\
7b & 6.4 & 24.8$\pm$0.6 \\
7c & 2.5 & 24.8$\pm$0.5 \\
8a & 5.6 & 23.5$\pm$0.4 \\
8b & 111.8 & 22.8$\pm$0.2 \\
9a & 3.6 & 25.3$\pm$0.4 \\
9b & 3.2 & 26.2$\pm$0.6 \\
9c & 4.5 & 27.4$\pm$0.9
\enddata

\end{deluxetable*}

\begin{deluxetable*}{lccc}
\centering
\tabletypesize{\scriptsize}
\tablecaption{Photometry for MACS0647-JD\label{t:delens_6}}
\tablewidth{0pt}
\tablehead{\colhead{} & \colhead{JD1} & \colhead{JD2} & \colhead{JD3}}
\startdata
F160W flux & 162$\pm$13nJy & 136$\pm$9nJy & 42$\pm$4nJy \\
$\mu$ & 7.9 & 7.7 & 2.6 \\
delensed F160W flux & 20.5$\pm$1.65nJy & 17.66$\pm$1.17nJy & 16.15$\pm$1.54nJy \\
$m_{AB}$ & 28.1$\pm$0.087 & 28.29$\pm$0.072 & 28.38$\pm$0.104 \\
$M_{UV}$ & -19.6 & -19.41 & -19.32
\enddata

\tablecomments{The F160W flux here was adapted from \citet{coe13} \\ Fluxes in nanoJanskys (nJy) may be converted to AB magnitudes via $m_{AB} \sim 26 - 2.5\log{(F_\nu/(145 nJy))}$}
\end{deluxetable*}

\begin{deluxetable*}{ccccc}
\centering
\tabletypesize{\scriptsize}
\tablecaption{Other Candidate Systems\label{table:zitrin_redshifts}}
\tablewidth{0pt}
\tablehead{\colhead{Image} & \colhead{RA} & \colhead{Dec} & \colhead{Phot-z [95\% C.I.]} & \colhead{Geo-z}}
\startdata
10.1 & 101.919490 & 70.249056 & 6.67 [6.27,7.28] & 7.3$\pm$0.7 \\
10.2 & 101.920490 & 70.244863 & 7.62 [1.05,8.03] & 7.3$\pm$0.7 \\
11.1 & 101.978410 & 70.253044 & 2.70 [2.47,3.03] & 1.95$\pm$0.1 \\
11.2 & 101.979890 & 70.249108 & ... & 1.95$\pm$0.1 \\
11.3 & 101.965730 & 70.240262 & 0.41 [0.12,0.53] & 1.95$\pm$0.1 \\
12.1 & 101.965030 & 70.246889 & ... & 2.3$\pm$0.1 \\
12.2 & 101.955950 & 70.242749 & 2.39 [1.83,2.45] & 2.3$\pm$0.1\\
12.3 & 101.967740 & 70.258397 & 2.39 [1.82,2.46] & 2.3$\pm$0.1\
\enddata

\tablecomments{The phtometric redshift of other candidate systems that are not included in mass model reconstruction and the predicted geometric redshifts from our model.}
\end{deluxetable*}

\newpage

\appendix

\section{Mass Map Uncertainties}

Although the mass model we reconstruct shows very high level of self-consistency and predicts well the relative brightnesses of lensed images, there is no strictly unique solution because the finite number of data points used effectively limits the spatial resolution of the reconstructed mass map. The grid construction is very general and hence inherently constrained in its freedom by the finite  number of Gaussian pixels adopted. Here we investigate the accuracy of the mass map and its sensitivity to initial guesses for finding the best fit due to the inherent assumptions adopted when reconstructing the mass model.

\subsection{Varying the Model Grid}

\begin{figure}
\centering
\includegraphics[width=85mm]{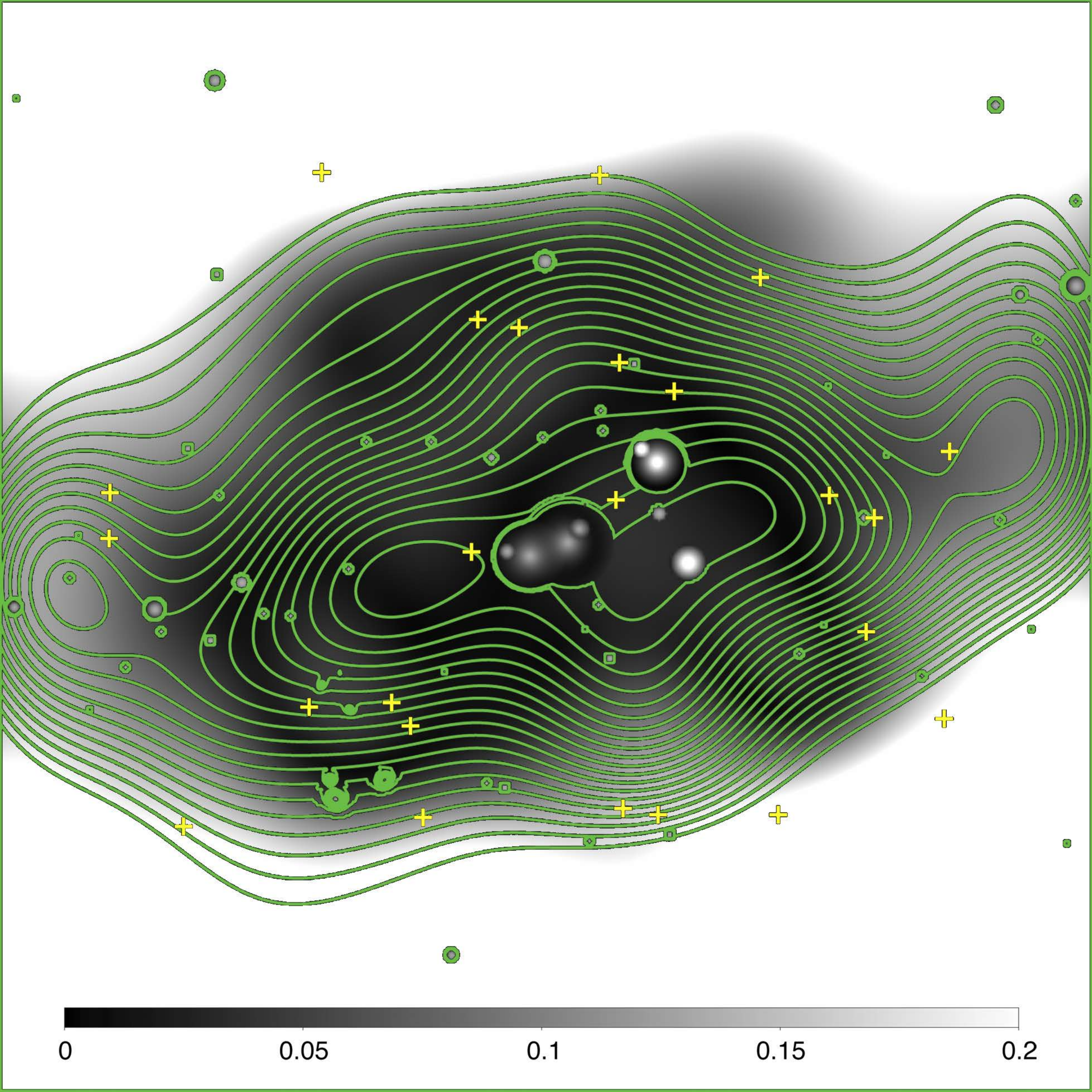}
\caption{
The background shows the ratio of the standard deviation map of our 16 mass models about the averaged mass map. These models differ by spanning a range of 3 in grid resolution. Mass contours of the mean mass model are overlaid on the mass variance map and the positions of multiply lensed images are overlaid on the map. Significant differences are only found outside the area covered by data.
}
\label{f:sd_grid}
\end{figure}

The reconstructed mass model will depend at some level on the number of Gaussian grids that we chose for the minimization. For our mass models derived above  we found that 32 $\times$ 32 = 1024 grid points was a suitable choice in which self-consistent lens models can be generated. Now we investigate other choices, by construction the mass model with a  relatively wide range of resolution: 23 $\times$ 23 = 529 to 38 $\times$ 38 = 1444 and measuring the standard deviation in the recovered mass over the surface of the mass map, as shown in figure \ref{f:sd_grid}. Here we can see  the central region is generally well constrained by the multiply lensed region, the variation is generally very small with $\sim$5\% differences. However, the variation is noticeably larger in the outer region where the model can not be as well constrained by the lack of distant counter images weel beyond the Einstein radius where the uncertainties rise to ~20\%.

\subsection{Initial Assumptions Sensitivity}

\begin{figure}
\centering
\includegraphics[width=85mm]{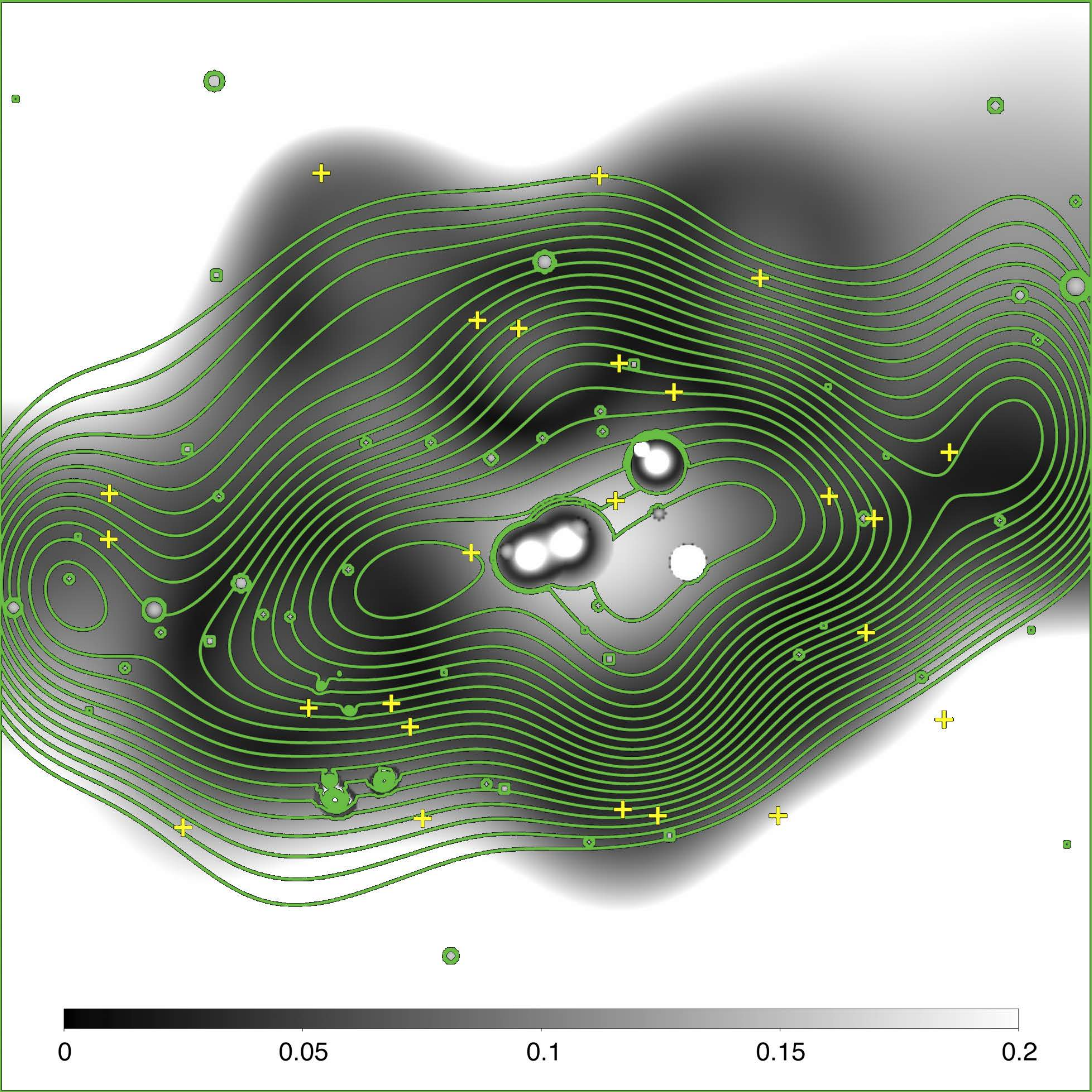}
\caption{
Same as figure \ref{f:sd_grid} but here shows the variability in initial guess.
}
\label{f:sd_initial}
\end{figure}

As described in section \ref{sec:model}, the mass model was determined by solving a system of linear equation (equation \ref{eq_lens_system}). The unknowns are solved by minimizing a quadratic function that estimates the solution of equations \ref{eq_lens_system}. The minimization procedure was carried out in a series of orthogonal conjugate directions with an initial guess for the solution. The initial guess include the source positions and the mass included in each grid cell. The algorithm stops at a certain value which is determined to be a local minimum around the initial guess. For a highly degenerate case like this we could reach to different local minimum points starting with different initial assumed values. We construct 15 different mass models by varying the initial assumption values with the range between 0.5 to 1.5 times the original value, with the uncertainties shown in Figure \ref{f:sd_initial}.

\subsection{Contribution of Photometric Redshift uncertainty to Mass Map Variance}

\begin{figure}
\centering
\includegraphics[width=85mm]{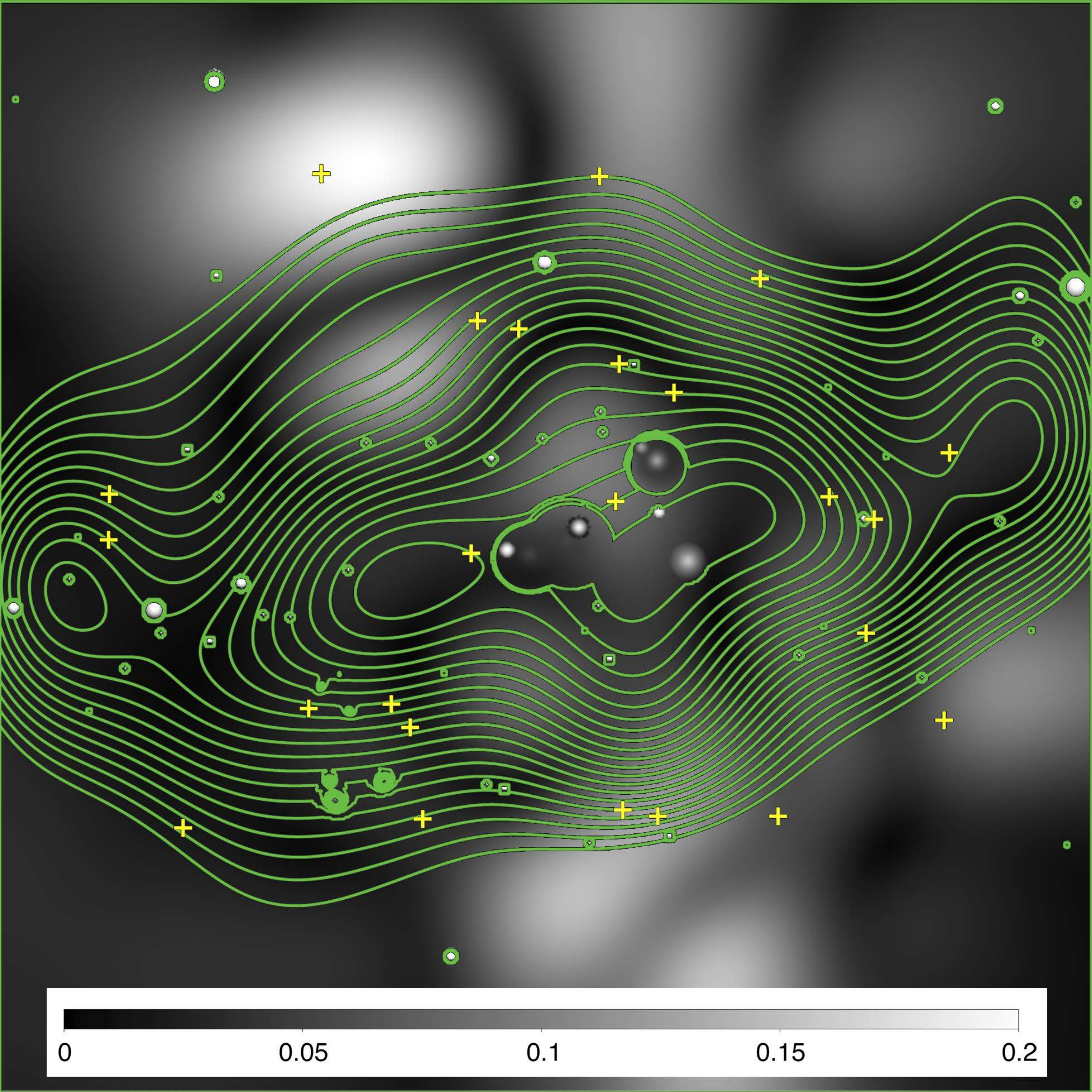}
\caption{
Same as figure \ref{f:sd_grid} but here shows the variability in Photometric redshift.
Very little uncertainty in the mass map is produced by randomly varying the photometric redshifts
within their respective estimated redshift uncertainties.
}
\label{f:sd_photoz}
\end{figure}

We also include the photometric redshift uncertainty in our mass model, as shown in Figure \ref{f:sd_photoz}. Assuming a Gaussian probability distribution centering at the photometric redshift with FWHM of the uncertainty predicted by \citet{coe13}. We reconstructed 15 mass models with the input redshifts of every source, drawn from a Gaussian distrubtion.

\subsection{Total Variance}

\begin{figure}
\centering
\includegraphics[width=85mm]{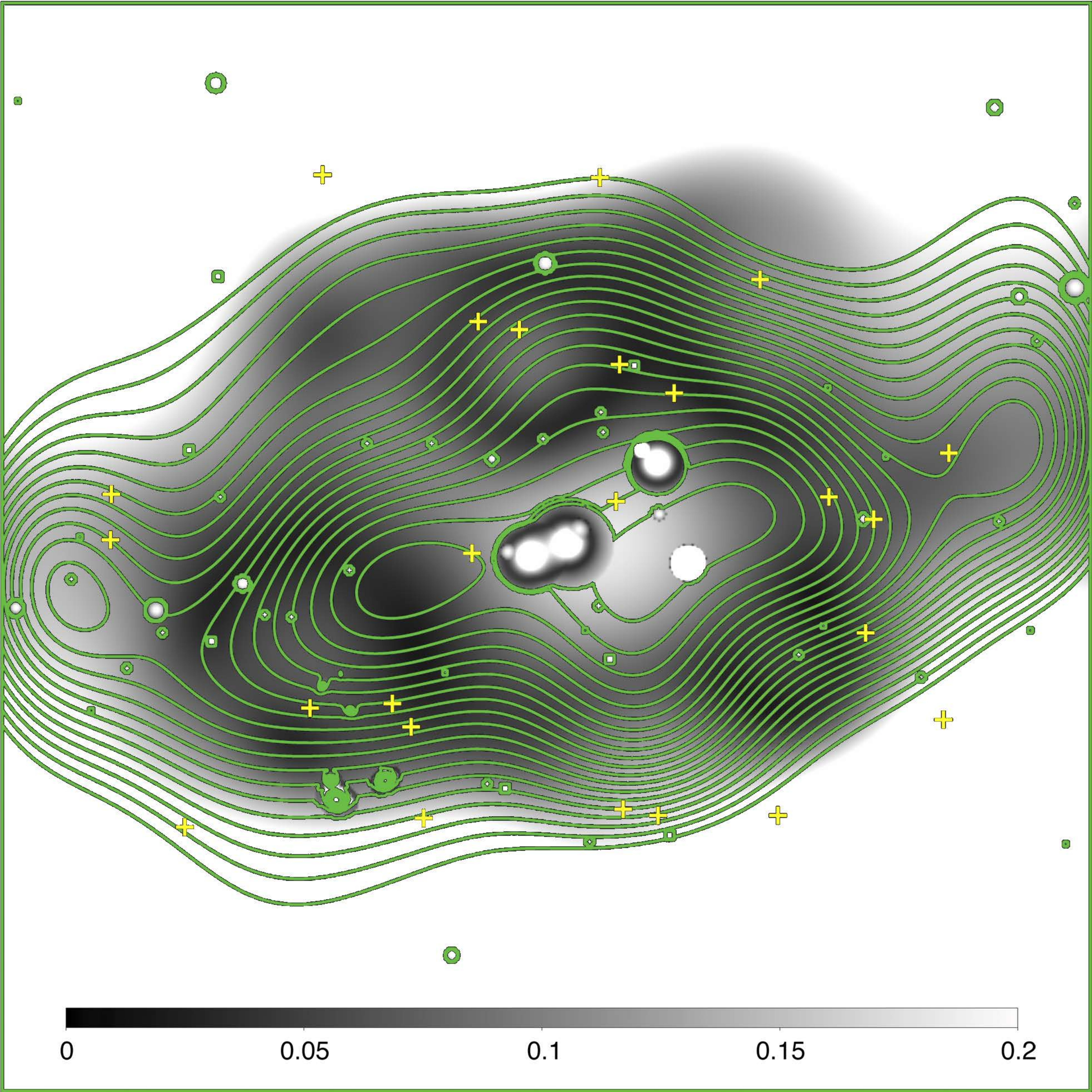}
\caption{
Same as figure \ref{f:sd_grid} but here shows the variability in all 3 effects. The dominant
effect comes from the choice of starting position for the minimisation, which for our purposes 
constitutes only a small uncertainty, that is noticeble near the center of mass at a level of 15\%. 
For most of the map only 5\% variance is attributable to all the above sources of noise for the region 
of interest, lying within the boundary of the multiply-lensed image distribution.
}
\label{f:sd_all}
\end{figure}

As shown in figure \ref{f:sd_all}, we conclude that in the center there is evidence that the dominant source of variance in the mass map 
comes from the choice of starting position for the minimisation, which for our purposes 
constitutes only a small uncertainty, that is noticeble within the central  $r<5\arcsec$ region, at a level of 15\%. For most of the map only 5\% variance is attributable to central region , lying within the boundary of the multiply-lensed images.

\end{document}